\documentclass{WileyMSP-template}

\usepackage{multirow}
\usepackage{multicol}
\usepackage{tablefootnote}

\usepackage{tabto}
\usepackage[table]{xcolor}
\usepackage{ragged2e}
\justifying
\usepackage{caption}
\usepackage{subcaption}
\usepackage{graphicx}
\usepackage[english]{babel}
\usepackage[latin1]{inputenc}
\usepackage{times}
\usepackage[T1]{fontenc}
\usepackage{inconsolata}
\usepackage{amsmath,bm}
\usepackage{amssymb}
\usepackage{hyperref}
\usepackage{color}
\usepackage{xcolor}
\usepackage{algorithm}
\usepackage{algpseudocode}
\usepackage{diagbox}
\usepackage{tabularx}
\usepackage{cuted}

\usepackage{booktabs}
\usepackage{tablefootnote}
\usepackage{array}
\usepackage{makecell}
\usepackage[export]{adjustbox}
\usepackage{parskip}
\usepackage{soul}
\usepackage{float}

\newcolumntype{L}[1]{>{\raggedright\arraybackslash}p{#1}}
\newcolumntype{C}[1]{>{\centering\arraybackslash}p{#1}}
\newcolumntype{R}[1]{>{\raggedleft\arraybackslash}p{#1}}
\begin{document}


\title{In-Situ Fault Detection of Submerged Pump Impellers Using

Encapsulated Accelerometers and Machine Learning}

\maketitle

\author{Sahil P. Wankhede}
\author{Xiangdong Xie}
\author{Ali H. Alshehri}
\author{Keith W Brashler}
\author{Mohammad Ba'adani}
\author{Doru C Turcan}
\author{Kamal Youcef-Toumi}
\author{Xian Du*}



\begin{affiliations}
Sahil P. Wankhede, Xian Du *\\
Department of Mechanical Engineering, University of Massachusetts Amherst, MA 01003, U.S.A
Center for Personalized Health Monitoring, (CPHM), Institute for Applied Life Sciences (IALS), MA 01003, U.S.A\\
Email Address:
xiandu@umass.edu

Xiangdong Xie\\ College of Information and Computer Science, University of Massachusetts Amherst, MA 01003, U.S.A 

Ali H. Alshehri,Keith W Brashler,Mohammad Ba'adani,Doru C Turcan.\\
Saudi Arabian Oil Company (Saudi Aramco), Dhahran 31311, Saudi Arabia

Kamal Youcef-Toumi\\
Massachusetts Institute of Technology, 77 Massachusetts Avenue, Cambridge, Massachusetts 02139, U.S.A.

\end{affiliations}



\begin{abstract}

\textcolor{black}{Vertical turbine pumps in oil and gas operations rely on motor-mounted accelerometers for condition monitoring. However, these sensors cannot detect faults at submerged impellers exposed to harsh downhole environments. We present the first study deploying encapsulated accelerometers mounted directly on submerged impeller bowls, enabling in-situ vibration monitoring. Using a lab-scale pump setup with 1-meter oil submergence, we collected vibration data under normal and simulated fault conditions. The data were analyzed using a suite of machine learning models---spanning traditional and deep learning methods---to evaluate sensor effectiveness. Impeller-mounted sensors achieved \textbf{91.3\% average accuracy} and \textbf{0.973 AUC-ROC}, outperforming the best non-submerged sensor. Crucially, encapsulation caused no statistically significant performance loss in sensor performance, confirming its viability for oil-submerged environments. While the lab setup used shallow submergence, real-world pump impellers operate up to hundreds of meters underground\textemdash well beyond the range of surface-mounted sensors. This 
in-situ monitoring system demonstrates that impeller-mounted sensors---encapsulated for protection while preserving diagnostic fidelity---can reliably detect faults in critical submerged pump components. By capturing localized vibration signatures that are undetectable from surface-mounted sensors, this approach enables earlier fault detection, reduces unplanned downtime, and optimizes maintenance for downhole systems in oil and gas operations.}


\end{abstract}


\section{Introduction}

\textcolor{black}{
Vertically suspended pumps are critical to oil and gas operations, enabling efficient fluid transfer across extensive extraction systems. Effective condition monitoring (CM) is essential to prevent unexpected failures---a single pump outage can cost millions of dollars in lost production. Traditional CM approaches rely on accelerometers mounted on above-ground motors\cite{corley1980vibration}; however, these are ineffective at detecting faults in the first-stage impeller---the primary submerged component responsible for initiating fluid displacement.}

\textcolor{black}{
The impeller operates in uniquely harsh conditions---submerged in crude oil at elevated temperatures (often >100$\degree$ C)\cite{lakal2022sensing}, with exposure to abrasive particulates and salinity gradients. This environment accelerates the degradation of conventional sensors \cite{corley1980vibration}, while routine maintenance is highly impractical due to long service intervals---typically up to two years\cite{lakal2022sensing}.}

\textcolor{black}{To address these challenges, directly mounting accelerometers on the first-stage impeller offers a promising solution for early fault detection. Building on our prior work demonstrating the viability of encapsulated sensors in high-temperature oil environments\cite{wankhede2022encapsulating, wankhede2023encapsulating,wankhede2023chem} (e.g., epoxy-fluoroelastomer composites resilient to 150\degree C oil immersion), we extend this approach by integrating encapsulated accelerometers onto the impeller bowls of vertical oil pumps---a critical advancement for in-situ condition monitoring.}

\begin{figure}[h]
    \centering
    \includegraphics[width=0.75\linewidth]{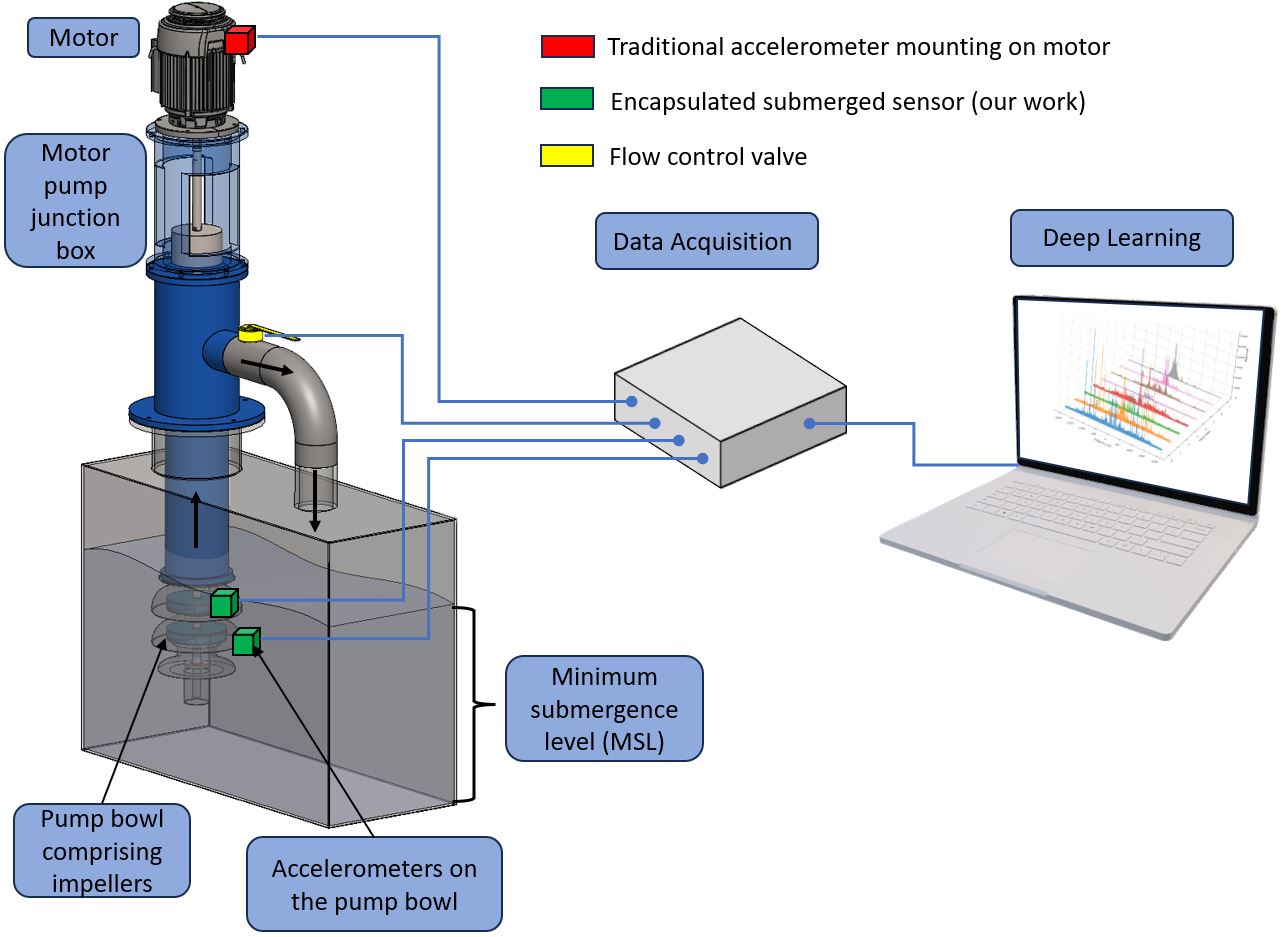}
    \caption{\centering Pictorial representation of the vertical pump system and overview of the research work . (In real oil field scenarios, the sections below the submergence level are significantly longer than those above it.)}
    \label{fig:pump_and_workflow}
\end{figure}

\textcolor{black}{Figure \ref{fig:pump_and_workflow} illustrates the vertically suspended pump assembly. Key above-ground components (motor, junction box, discharge outlet) contrast with the submerged impeller bowl assembly connected via a flanged column. The manufacturer-specified minimum submergence level (MSL) is annotated, highlighting the operational constraints that motivate our impeller-mounted sensor deployment.}

\textcolor{black}{Traditional condition monitoring for oil pumps typically involves placing accelerometers on the above-ground motors. In contrast, this study proposes direct sensor installation onto the oil-submerged impeller---enabled by the novel sensor encapsulation. Previous studies have extensively investigated vibration monitoring in horizontal centrifugal pumps (e.g. Zhang et al.\cite{zhang2022vibration}, Bai et al.\cite{bai2019vibration}, Wang et al. \cite{Wang2018} ), which facilitate easy sensor mounting; however, vertical oil pumps pose unique challenges. As shown in Table \ref{tab:pump_literature}, most existing work uses above-ground sensors, with only 2 of 27 studies reporting impeller-bowl measurements. The submerged impeller environment necessitates encapsulation-protected sensors capable of withstanding oil exposure, temperature extremes, and complex installation constraints.}

\newcolumntype{Y}{>{\RaggedRight\arraybackslash}X}

\begin{table*}[t]
\caption{Classification of Pump Studies}
\label{tab:pump_literature}
\centering
\begin{minipage}{1\linewidth}
\renewcommand{\thempfootnote}{\arabic{mpfootnote}}
\renewcommand{\arraystretch}{1.5}
\setlength{\tabcolsep}{5pt}
\begin{tabularx}{\textwidth}{L{3cm}L{2cm}L{2.5cm}L{4cm}C{3cm}C{1.2cm}}
\toprule[1.5pt]
\textbf{References} &\textbf{Pump Type}&\textbf{Sensors}&\textbf{Mounting location}&\textbf{Fluid}	\textbf{Sensor}&\textbf{Submerged in fluid}\\
\midrule[1pt]
Leader et al.\cite{Leader1985}\newline 1985 & Vertical Pump & Velocity\newline transducers & Motor & Toxic liquid in chemical plant & No 
\\
\midrule[0.1pt]
Smith et al.\cite{Smith1986}\newline 1986 & Vertical Pump & Velocity\newline transducers & Near the impeller & Water & Yes 
\\
\midrule[0.1pt]
Schiavello et al.\cite{Schiavello2004}\newline 2004 & Vertical Pump & Accelerometers \newline\&\newline proximity probe & Motor and pump inlet bell, proximity probe near the impeller & Water & No 
\\
\midrule[0.1pt]
Thuyetle et al.\cite{Le_TurbinePump}\newline 2008 & Vertical Pump & Vibration\newline transducer & Motor, bearing, discharge pipe, and mechanical seal & Oil & No 
\\
\midrule[0.1pt]
Shahrooi et al.\cite{Shahrooi2009}\newline 2009 & Vertical Pump & Velocity\newline transducers & Foundation, upper bearing \newline
\& \newline
lower bearing of the motor & Water & No 
\\
\midrule[0.1pt]
Abdel-Rahman et al.\cite{AbdelRahman2009}\newline 2009 & Vertical Pump & Accelerometers & Motor, pump, bearings, and foundation & Water & No 
\\
\midrule[0.1pt]
Leader et al.\cite{LeaderConnerLucas}\newline 2010 & Vertical Pump & Velocity\newline transducers & Motor & Liquid sulfur & No 
\\
\midrule[0.1pt]
Fetyan et al.\cite{Fetyan2014}\newline 2014 & Vertical Pump & Accelerometers & Motor, motor base plates, pump bearing & Water & No 
\\
\midrule[0.1pt]
El-Gazzar et al.\cite{ElGazzar2017}\newline 2017 & Vertical Pump & Accelerometers & Motor upper and lower bearings and pump bearings & Water & No 
\\
\midrule[0.1pt]
Stan et al.\cite{Stan2018}\newline 2018 & Horizontal centrifugal pump & Accelerometers & Pump housing & Water & No 
\\
\bottomrule[1.5pt]
\end{tabularx}
\end{minipage}
\end{table*}

\textcolor{black}{Condition Monitoring (CM) of pumps is a crucial aspect of industrial maintenance, aimed at detecting potential failures at their earliest stages. This proactive approach minimizes non-productive time, prevents catastrophic failures, and optimizes maintenance schedules. Numerous studies have advanced robust CM systems, focusing on real-time sensor data analysis. Similarly, Predictive Maintenance (PM) leverages data-driven models to predict failures and estimate component lifespans. By identifying subtle deviations in operational patterns, PM enhances reliability through early fault detection.}


\textcolor{black}{Both CM and PM systems leverage real-time sensor data to detect operational deviations through anomaly detection algorithms. These systems first establish baselines of normal vibration patterns (e.g., RMS values, spectral profiles) and then flag statistical outliers in feature space. While machine learning has become instrumental in processing these high-dimensional signals, vibration data from industrial equipment like oil pumps presents unique challenges: high sampling rates (4 kHz in this study), transient fault signatures, and signal contamination from adjacent machinery.}

\textcolor{black}{To overcome these challenges, feature extraction techniques distill raw signals into discriminative metrics-first-order statistics (mean, RMS), second-order moments (variance, kurtosis), and spectral descriptors. These features enable simpler anomaly detection models to match or exceed deep learning performance in many industrial applications\cite{AlTobi2019}, particularly when fault signatures manifest in specific frequency bands or statistical distributions. Our work extends this paradigm to submerged impeller monitoring, where encapsulation-induced signal modifications may require more careful feature engineering.}

\begin{figure}
    \centering
    \includegraphics[width=1\linewidth]{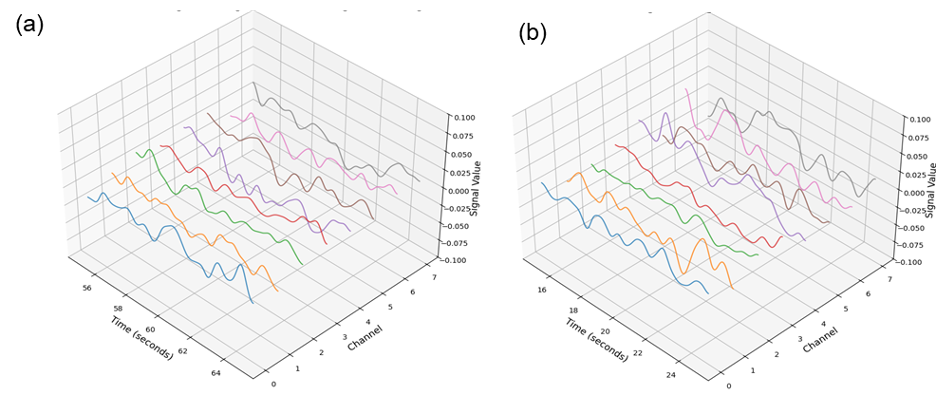}
    \caption{\centering The Gaussian blurred signals from 8 sensors. \protect\linebreak 
    (a) Gaussian Blurred signals of eight channels during valve change (b) Gaussian Blurred signals of eight channels during normal flow. (Note: Signals are Gaussian blurred with sigma = 200 as the sampling frequency 4000 is too high for visualization.)}
    \label{fig:gaussian_blurred_signals}
\end{figure}

\textcolor{black}{As shown in Figure \ref{fig:gaussian_blurred_signals} , detecting pump condition changes is non-trivial, as there are minimal visually discernible signal pattern differences---necessitating machine learning approaches. Given the time-series nature of vibration data and our objective---evaluating impeller sensors' diagnostic capability---we employ a diverse set of ML models to comprehensively assess anomaly detectability.}

\textcolor{black}{Guided by recent anomaly detection benchmarks\cite{ADBench,revisitingTSOD} and the no-free-lunch theorem\cite{noFreeLunchTheorem}, we hypothesize that distinguishable anomalies will be detected by some (but not all) models. Accordingly, we selected 10 models spanning distinct methodologies: classical feature-based (e.g., Isolation Forest, KNN), deep learning (e.g., LSTM-Autoencoder, Anomaly-Transformer), parametric vs. non-parametric, and time-series-specialized vs. tabular-data-optimized architectures\cite{Lai2021RevisitingTSOD}. Each model underwent hyperparameter tuning across 16 feature/window configurations to evaluate the informativeness of the sensor data .}

\textcolor{black}{Furthermore, considering the difficulty of obtaining anomalous data from operational oil pumps---as physical anomaly simulation is both technically challenging and potentially wasteful---we employed unsupervised and semi-supervised models trained exclusively on normal operation data. The selected models include:} 

\textcolor{black}{\begin{itemize}
    \item \textbf{Six traditional feature-based models}\cite{Lai2021RevisitingTSOD}: Isolation Forest\cite{Liu2008IForest} (iForest), K-Nearest Neighbors\cite{Fix1989(KNN)} (KNN), Clustering-Based Local Outlier Factor\cite{He2003} (CBLOF), Copula-Based Outlier Detection\cite{Li2020COPOD} (COPOD),\\ Autoencoder\cite{Rumelhart1986}, and Deep Support Vector Data Description\cite{Ruff2018} (DeepSVDD)
    \item \textbf{Four time-series-specific deep learning models}: LSTM-Autoencoder\cite{Malhotra2016} (LSTM-AE, competitive since 2016\cite{wagner2023timesead, Schmidl2022}), Temporal Convolutional Network for Anomaly Detection\cite{HeZhao2019} (TCN-AD), Temporal Hierarchical One-Class Network\cite{Shen2020} (THOC), and Anomaly-Transformer\cite{Xu2022}
\end{itemize}}

\textcolor{black}{These deep learning models leverage sequential patterns through architectures like self-attention and dilated convolutions, frequently serving as baselines in time-series anomaly detection literature\cite{Kim2022,ZamanzadehDarban2024,Choi2021,Zhang2024}. Each of the aforementioned models is briefly introduced along with its key strengths for anomaly detection below:}

\begin{itemize}
    \item \textit{Isolation Forest (iForest)}: iForest detects anomalies by randomly selecting a feature and recursively partitioning the data along its range. Anomalies, being fewer and distinct from normal data points, require fewer splits to isolate. iForest is highly efficient, especially for high-dimensional datasets, as it operates in linear time relative to the number of samples. It has scalability to large datasets, minimal assumptions about data distribution, and the ability to handle both univariate and multivariate anomalies effectively.
    \item \textit{Autoencoder (AE)}: AE reduces high-dimensional time series data into a low-dimensional latent space, retaining key patterns for reconstruction. During training, AE learns to minimize reconstruction errors for normal data patterns. When applied to unseen data, anomalies that deviate significantly from the learned normal patterns exhibit higher reconstruction errors. AE is well known for its flexibility in handling complex, high-dimensional data and its ability to detect subtle deviations by capturing non-linear relationships within the data.
    \item \textit{K-Nearest Neighbors (KNN)}: KNN identifies anomalies by measuring the distance between a data point and its nearest neighbors. Data points that lie far from their neighbors are deemed anomalous. KNN is a non-parametric approach, effective when normal data forms clusters, as anomalies stand out by proximity.
    \item \textit{Clustering-Based Local Outlier Factor (CBLOF)}: CBLOF combines clustering with local outlier detection. It assigns each data point a score based on its cluster size and distance from the cluster center, identifying points in small or distant clusters as anomalies. Hence, CBLOF has the ability to handle both global and local anomalies, adaptability to diverse data distribution, and effectiveness in identifying context-dependent outliers.
    \item \textit{Copula-Based Outlier Detection (COPOD)}: COPOD computes the empirical copula distribution for each data point and assigns an anomaly score based on the deviation from expected patterns. COPOD has the scalability to large datasets, interpretability due to its statistical foundation, and its ability to handle both univariate and multivariate anomalies effectively. 
    \item \textit{One-Class Support Vector Machine (SVM)}: This model defines a decision boundary around the normal data points in feature space and flags data points that fall outside this boundary as anomalies. One-Class SVM has the ability to model non-linear decision boundaries using kernel functions, robustness to high-dimensional data, and suitability for cases where only normal data is available for training
    \item \textit{LSTM-Autoencoder (LSTM-AE)}: Leveraging an encoder-decoder structure with LSTM units, LSTM-AE reconstructs time series sequences. By minimizing reconstruction errors during training on normal data, the model learns typical temporal patterns.  Anomalies, which deviate from these patterns, result in higher reconstruction errors. LSTM-AE can handle complex temporal dependencies, detect subtle anomalies in sequential data, and adapt to high-dimensional time-series datasets.
    \item \textit{Temporal Convolutional Network (TCN)}: TCN leverages 1D dilated causal convolutions and residual connections to capture long-range temporal dependencies while maintaining sequential order efficiently. It is effective for anomaly detection by capturing both local and global temporal patterns while maintaining computational efficiency ensured by its parallelizable architecture. 
    \item \textit{Anomaly-Transformer}: The Anomaly-Transformer leverages a self-attention mechanism of Transformers to capture local and global dependencies. A novel "Anomaly-Attention" module, combined with a distribution modeling approach, enables it to focus on critical time steps and detect subtle and context-dependent anomalies in temporal data. Anomaly-Transformer can handle complex temporal structures and long sequences while robustly detecting anomalies with varying characteristics across diverse datasets.
    \item \textit{Temporal Hierarchical One-Class Network (THOC)}: THOC leverages a hierarchical structure to learn temporal dependencies at multiple resolutions, capturing both short-term patterns and long-term trends in the time series data. It employs one-class classification principles at each level to detect anomalies by identifying deviations from the learned normal patterns across hierarchical temporal scales. Hence, THOC can detect anomalies across different temporal scales, and adapt to diverse time-series datasets.

\end{itemize}

\textcolor{black}{This selection provides complementary capabilities: traditional feature-based models offer computational efficiency and interpretability, while deep learning architectures capture complex temporal patterns through automated feature learning.}

\textcolor{black}{In this study, we deployed encapsulated accelerometers on the impeller bowls of a vertical oil pump to enable early-stage failure detection. Anomalous conditions were simulated by modulating flow rates and oil levels, with vibration data collected from impeller-, motor-, and bearing-mounted sensors. Results demonstrate that impeller-mounted sensors provide superior diagnostic information---achieving 91.3\% average fault detection accuracy---due to their proximity to fault origins. The encapsulation preserved sensor functionality in harsh oil-immersed environments, with only slight performance degradation compared to non-encapsulated sensors.We trained 10 ML models on this dataset, establishing a mini-benchmark for submerged sensor efficacy in pump monitoring. To our knowledge, this is the first intelligent in-situ impeller-mounted sensing system---encapsulated for protection while maintaining diagnostic fidelity---that can reliably detect faults in critical submerged pump components. Figure \ref{fig:pump_and_workflow} summarizes the experimental workflow, with detailed methodologies in subsequent sections.}

Our key contributions are:
\textcolor{black}{
\begin{enumerate}
    \item We present the first intelligent in-situ monitoring system using encapsulated, impeller-mounted accelerometers for fault detection in oil pumps, achieving a 5.3\% accuracy improvement over surface-mounted sensors under harsh downhole conditions.
    \item We demonstrate that some features extracted from sliding windows significantly enhance anomaly detection, whereas raw vibration signals are insufficient.
    \item To advance benchmarking and reproducibility, we open-source a comprehensive dataset and codebase encompassing 16 feature/window configurations and 10 ML models.
\end{enumerate}
}

\section{Results and Discussion}

We analyze the experimental results from three perspectives: 1) the informativeness of the impeller's bowl sensor compared to other sensors, 2), the influence of hyperparameters on model performance, and 3), the impact of encapsulation on the impeller's bowl sensor.

\textcolor{black}{We evaluated the diagnostic capability of impeller-mounted accelerometers from three key perspectives: (1) comparative sensor informativeness, (2) the impact of encapsulation on signal fidelity, and (3) the influence of hyperparameters on detection performance. ML models were trained on vibration data from eight sensors---seven above-oil and one encapsulated impeller-mounted---under both normal and simulated fault conditions, including valve closures and abnormal oil-level changes.}

\textcolor{black}{Data preprocessing included signal standardization, sliding window segmentation (0.25--3s durations), and feature extraction for feature-based models, including 13 statistical and 15 spectral features per sensor.  For deep models, we applied a Gaussian kernel moving average with step sizes ranging from 0.025 to 0.125s to reduce data length.  Model robustness was evaluated using five-fold cross-validation (CV): normal data was split into 5 parts, and in each fold, one part of the normal data along with all the anomalous data served as the test set, while the remaining normal data was used for training. Performance was accessed using AUC-ROC and average accuracy, where accuracy was computed based on 5 different thresholds derived from percentiles of anomaly scores of the training data (see Section III for experimental details). The following sections present:}

\begin{itemize}
    \item \textbf{Sensor Informativeness}: Impeller vs. motor/bearing sensors
    \item \textbf{Encapsulation Impact}: Protected vs. bare impeller sensors  
    \item \textbf{Feature/Parameter Effects}: Statistical vs. spectral features, window configurations
\end{itemize}

\subsection{Results}

\subsubsection{Cross-Validation and General Observations}

\begin{figure}[h]
    \centering
    \includegraphics[width=.95\linewidth, height=.5\linewidth]{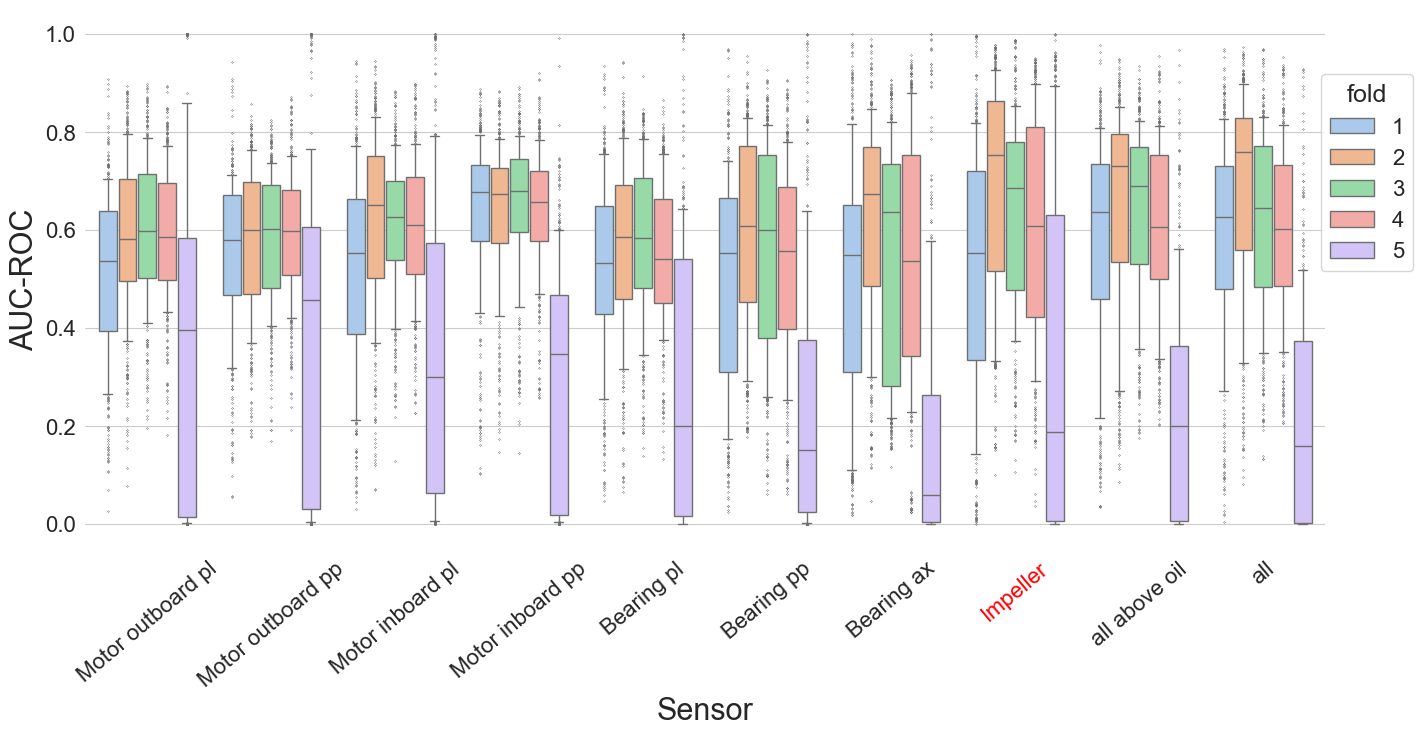}
    \caption{\centering  AUC-ROC of all ML models on data from sensors in each fold. \protect\linebreak Small dots indicate outliers below 10\% or above 90\% percentile.}
    \label{fig:AUCROC_fold_sensor}
\end{figure}

\textcolor{black}{Five-fold CV revealed substantial performance variability across data splits (Figure \ref{fig:AUCROC_fold_sensor}), with AUC-ROC scores fluctuating by up to 0.6 between folds. Models achieved peak performance in fold 2, which used normal data from 75\% abrupt valve closures as the normal testing data, while fold 5---containing constant flow data as the normal testing data---showed remarkably lower detection capability. This variability suggests that certain flow signals are more representative and some show more similarities with anomalous signals. While the exact root cause of Fold 5's performance degradation remains unclear, we evaluated both mean and maximum performance across folds for each sensor and hyperparameter configuration. As only Fold 5 showed a pronounced drop in AUC-ROC, we suspect an unknown issue in that fold data. Given that anomalous data only appears in the test set, and assuming consistency across normal data, the maximum AUC-ROC across folds provides a better indication of the diagnostic value learned from training data. Therefore, we mainly report maximum AUC-ROC across folds for each configuration as the baseline performance. For brevity, we refer to this maximum value as "AUC-ROC" in the remainder of the paper. The same convention applies to average accuracy.}

\textcolor{black}{Deep learning models achieved marginally higher peak AUC-ROC scores than traditional ML models, but exhibited significantly more sensitivity to hyperparameter choices (Table \ref{tab:default_hyperparameters}) and data variability. In contrast, untuned traditional models using default parameters demonstrated greater stability across conditions. This robustness---combined with substantially faster training times---makes traditional models such as Isolation Forest, COPOD practically advantageous for real-time pump monitoring, despite slightly lower maximum performance.}

\textcolor{black}{We focus our analysis on AUC-ROC and average accuracy, as these metrics effectively account for the class imbalance in our testing data.}

\subsubsection{Sensor Informativeness}

\textcolor{black}{Encapsulated and non-encapsulated impeller sensors significantly outperformed all other sensor configurations, achieving a \textbf{median maximum AUC-ROC of 0.950} for feature-based models. This surpassed combined above-oil sensors (0.911) and the top-performing non-impeller sensor (bearing axial: 0.901).}

\textcolor{black}{Deep learning models showed less variation in performance across sensor locations; however, impeller-mounted sensors still achieved the highest maximum AUC-ROC across CV folds. Interestingly, three motor sensors marginally outperformed the impeller sensor for LSTM-Autoencoder and Anomaly-Transformer models---an artifacts of the lab setup where sensor proximity (<1m) may have artificially enhanced their signal fidelity. Such proximity is unrealistic in real-world oilfield deployments, where impellers are typically located hundreds of meters below the surface.}

\textcolor{black}{These results confirm the impeller's unique ability to capture localized fault signatures, particularly for feature-based models. While lab conditions allow above-oil sensors to perform somehow comparably due to close proximity, this advantage would not translate to real oilfield settings, where impeller operate hundreds of meters below the surface. At such real distances, surface-mounted sensors are too far from the fault source to be effective, underscoring the necessity of in-situ encapsulated monitoring.}

\begin{figure}[h]
    \centering
    \includegraphics[width=1\linewidth]{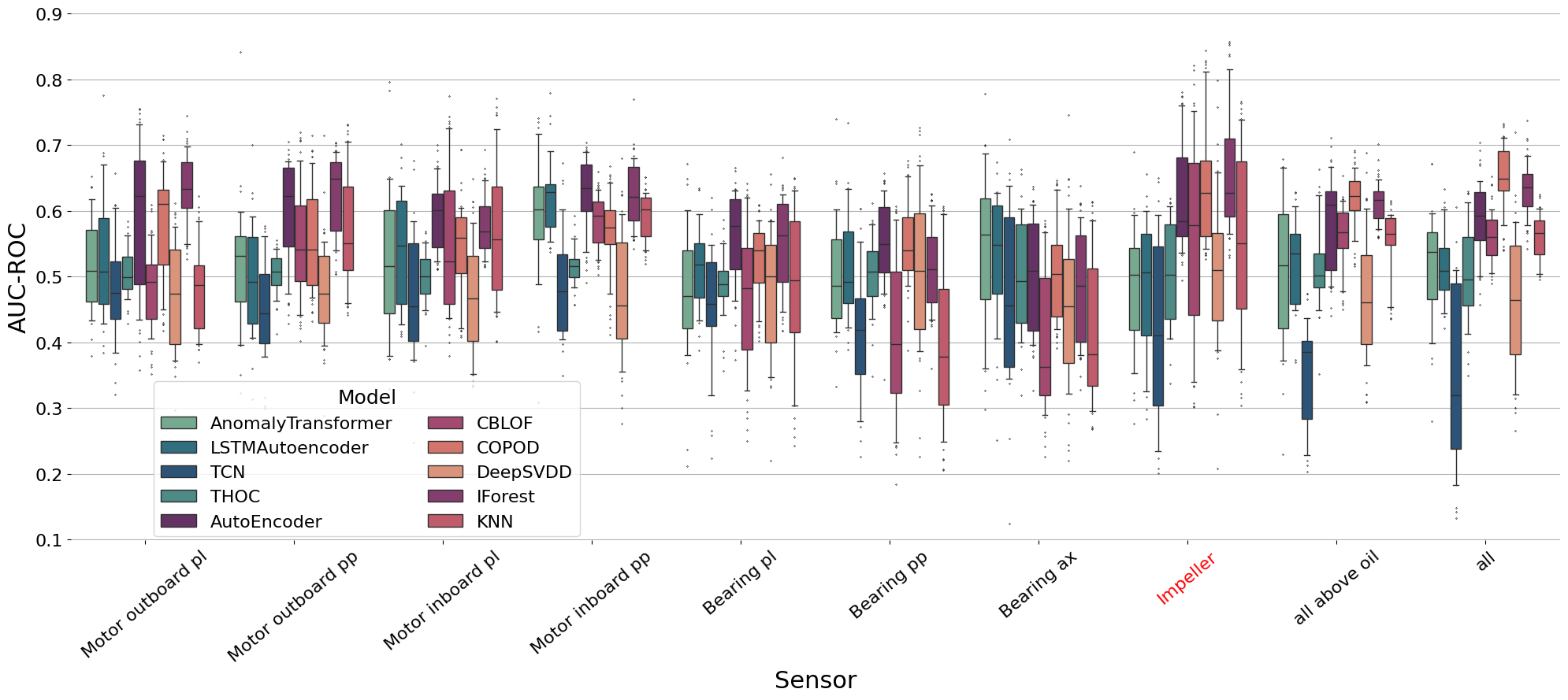}
    \caption{\centering \textbf{Mean} AUC-ROC across CV folds of ML models on data from each sensor. \protect\linebreak Deep models are colored with cold palettes and traditional models with warm palettes.}
    \label{fig:AUCROC_sensor_model}
\end{figure}

\subsubsection{Impact of Encapsulation}

\newcolumntype{Y}{>{\RaggedRight\arraybackslash}X}

\begin{table*}[t]
\caption{Accuracies of non-encapsulated sensor at different directions vs. encapsulated sensor.}
\label{tab:sensor_accuracy_directions}
\centering
\begin{minipage}{1\linewidth}
\renewcommand{\thempfootnote}{\arabic{mpfootnote}}
\renewcommand{\arraystretch}{1.3}
\setlength{\tabcolsep}{5pt}
\begin{tabularx}{\textwidth}{L{5.5cm}YYYY}
\toprule
\textbf{Sensor} & \textbf{Mean AUC-ROC} & \textbf{Mean Average Accuracy} & \textbf{Max AUC-ROC} & \textbf{Max Average Accuracy} \\
\midrule
encapsulated accelerometer (Sr\# 8a)                              & 81.6\% & 63.0\% & 97.3\% & 91.3\% \\
non-encapsulated accelerometers in parallel direction (Sr\# 8b)     & 82.4\% & 65.9\% & 99.7\% & 98.9\% \\
non-encapsulated accelerometers in perpendicular direction (Sr\# 8c) & 81.9\% & 66.5\% & 99.4\% & 98.0\% \\
\bottomrule
\end{tabularx}
\end{minipage}
\end{table*}

\textcolor{black}{To evaluate the effect of  encapsulation, we compared the encapsulated impeller-mounted sensor (Sr\#8a) with two non-encapsulated counterparts mounted in parallel (Sr\#8b) and perpendicular (Sr\#8c) directions. The strong agreement between Sr\#8a and Sr\#8b (Table~\ref{tab:sensor_accuracy_directions}), as well as between Sr\# 8b and Sr\# 8, indicated that dual-axis measurements may be redundant. Consequently, only the encapsulated sensor (Sr\#8a) was deployed in subsequent experiments to minimize installation complexity and other complications.}

To quantify the impact, we first examine the maximum AUC-ROC scores across CV folds of all models using data from the impeller sensor with and without encapsulation, by calculating the mean and standard deviation, checking the normality of the score distribution, performing Mann-Whitney U test, and computing the 95\% confidence interval (CI). Then, we selected 5 top performing models and did the testing again, based on each model's AUC-ROC scores which are demonstrated in Table \ref{tabS:AUCROC_sensor_model} in the supplementary. The final testing results are shown in Table \ref{tab:impact_of_encapsulation}. When considering all models, there is almost no difference between encapsulated and non-encapsulated sensors. However, looking at the top performing models, some exhibit moderate AUC-ROC reduction with encapsulation, while others exhibit huge AUC-ROC improvement with encapsulation. It is very interesting that models display opposite behaviors in detection rate on the sensor encapsulation. This implies that while the encapsulation at the impeller accelerometer changes what it detects, the information needed for fault detection stays, and no model is good at capturing every feature of this information. Because the best fault detection performances with and without encapsulation are the same, and the distribution of all model performances remains the same, we conclude that encapsulation preserves critical fault signatures while enabling reliable sensor deployment in oil-immersed conditions.

\newcolumntype{Y}{>{\Centering\arraybackslash}X}
\newcolumntype{M}[1]{>{\RaggedRight\arraybackslash}m{#1}}

\begin{table*}[h]
\caption{Statistical Analysis of Model Performances with Data from Encapsulated vs Non-encapsulated Impeller Sensor}
\label{tab:impact_of_encapsulation}
\centering
\begin{minipage}{1\linewidth}
\renewcommand{\thempfootnote}{\arabic{mpfootnote}}
\renewcommand{\arraystretch}{1.6}
\setlength{\tabcolsep}{5pt}
\begin{tabularx}{\textwidth}{M{2.5cm}YYYYYY}
\toprule

\multirow{1.6}{*}{\textbf{Comparison}} &  \multirow{1.6}{*}{\textbf{$\mu$}} &  \multirow{1.6}{*}{\textbf{$\sigma$}} & \textbf{Shapiro-Wilk Normality} & \textbf{Mann-Whitney U test} & \multirow{1.6}{*}{\textbf{95\% CI}}& \multirow{1.6}{*}{\textbf{Effect Size}}\footnote{We used Cohen'sd effect size with mean absolute deviation (MAD) normalization as the AUC-ROC distribution is highly non-Gaussian} \\

\midrule

\multirow{2}{*}{All Models} & 0.836$^*$ \footnote{Every first row in a cell is marked by $*$ corresponding to the encapsulated sensor data. The second row corresponds to the non-encapsulated one.} & 0.099 & 0 &  
\multirow{2}{*}{\shortstack{U=352704\\ p=0.501}}  & $0.830-0.843$ & \multirow{2}{*}{0.022} 
\\
\cmidrule{2-4} \cmidrule{6-6}
& 0.830 & 0.115 & 0 &  &  $0.822-0.838$ & \\

\midrule

\multirow{2}{*}{IForest} & 0.869$^*$ & 0.067 & 0 &  
\multirow{2}{*}{\shortstack{U=2592\\ p=0}}  & $0.855-0.882$ & \multirow{2}{*}{-0.275} 
\\
\cmidrule{2-4} \cmidrule{6-6}
& 0.919 & 0.036 & 0 &  &  $0.912-0.926$ & \\ 

\midrule

\multirow{2}{*}{COPOD} & 0.822$^*$ & 0.050 & 0 &  
\multirow{2}{*}{\shortstack{U=3232\\ p=0}}  & $0.811-0.832$ & \multirow{2}{*}{-0.247} 
\\
\cmidrule{2-4} \cmidrule{6-6}
& 0.884 & 0.099  & 0 &  &  $0.864-0.904$ & \\ 

\midrule

\multirow{2}{*}{CBLOF} & 0.910$^*$ & 0.048  & 0 &  
\multirow{2}{*}{\shortstack{U=8896\\ p=0}}  & $0.901-0.920$ & \multirow{2}{*}{0.823} 
\\
\cmidrule{2-4} \cmidrule{6-6}
& 0.759 & 0.058 & 0 &  &  $0.747-0.771$ & \\ 

\midrule

\multirow{2}{*}{KNN} & 0.897$^*$ & 0.072  & 0 &  
\multirow{2}{*}{\shortstack{U=8112\\ p=0}}  & $0.883-0.912$ & \multirow{2}{*}{0.605} 
\\
\cmidrule{2-4} \cmidrule{6-6}
& 0.780 & 0.056 & 0 &  &  $0.769-0.792$ & \\ 

\midrule

\multirow{2.5}{*}{\shortstack{Anomaly\\Transformer}} & 0.846$^*$ & 0.064  & 0 &  
\multirow{2}{*}{\shortstack{U=8112\\ p=0}}  & $0.883-0.912$ & \multirow{2}{*}{-0.302} 
\\
\cmidrule{2-4} \cmidrule{6-6}
& 0.895 & 0.074 & 0 &  &  $0.769-0.792$ & \\ 

\bottomrule
\end{tabularx}
\end{minipage}
\end{table*}

\subsubsection{Impact of Feature Extractions}

\begin{figure}
    \centering
    \includegraphics[width=1\linewidth]{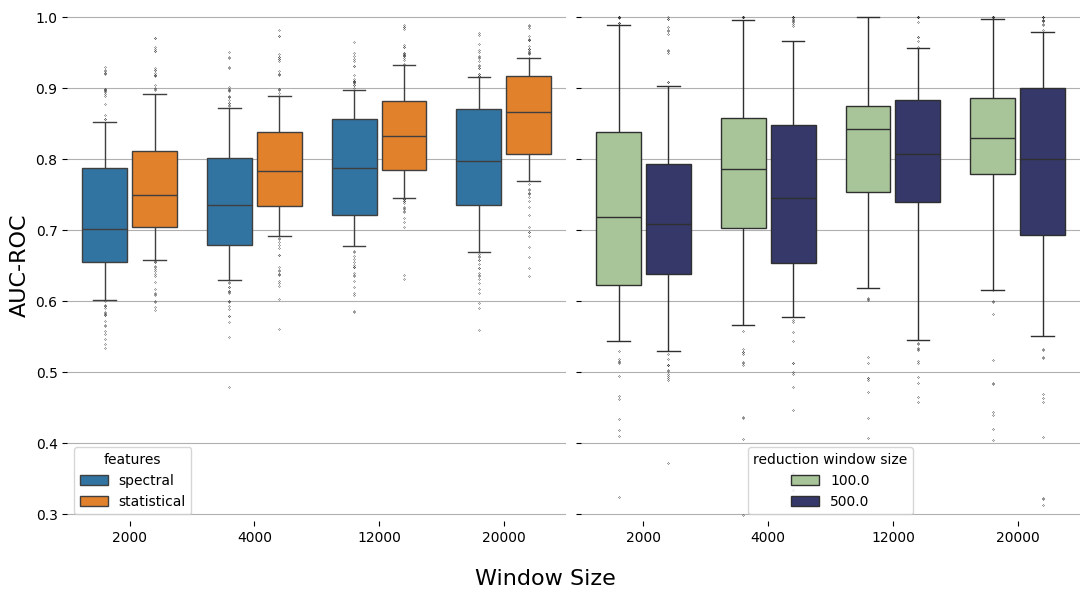}
    \caption{\centering AUC-ROC of top performing models on different window sizes and features. Left: results from IForest, COPOD, and AutoEncoder; Right: results from AnomalyTransformer, LSTMAutoencoder, and TCN.}
    \label{fig:AUCROC_features}
\end{figure}

\textcolor{black}{Figure \ref{fig:AUCROC_features} demonstrates that statistical features consistently outperform spectral features across models, and that longer sampling windows lead to better anomaly detection performance. Deep learning models generally achieved improved accuracy with extended windows, although their robustness remained relatively unaffected by shorter window size.}


\textcolor{black}{Model performance remained consistent across sliding window overlaps (0--90\%) and positions, indicating robustness to segmentation artifacts. This suggests that once a sufficient number of windows are sampled, precise temporal alignment has minimal impact on the accuracy of trained model.}

\textcolor{black}{Notably, in some settings, non-submerged sensors achieved comparable performance to the impeller sensor in this lab environment (1m submergence). However, in real oilfield deployments where impellers are hundreds of meters submerged, vibration attenuation would render surface sensors ineffective---highlighting the necessity of impeller-mounted monitoring.}

\subsubsection{Threshold and Accuracy}

Surprisingly, we notice that the average accuracies obtained from the feature-based models are significantly worse than those from the deep models, while feature-based models provide higher AUC-ROC---average accuracies of deep models are shown in Table \ref{tab:encapsulated_thresholds}, and traditional feature-based models show at least 10\% lower accuracies at every threshold. This suggests that although different models can have similar performance when an ideal anomaly threshold is selected, some models learn the normal data with a more "reasonable" threshold or boundaries that can be applied more easily to unseen data in the common range from 0.1\% to 20\% for anomaly detection.

\begin{table*}[t]
\caption{Accuracies of encapsulated sensor at different thresholds (deep models).}
\label{tab:encapsulated_thresholds}
\centering
\begin{minipage}{1\linewidth}
\renewcommand{\thempfootnote}{\arabic{mpfootnote}}
\renewcommand{\arraystretch}{1.3}
\setlength{\tabcolsep}{5pt}
\begin{tabularx}{\textwidth}{L{4cm}C{4cm}C{4cm}C{4cm}}
\toprule
\textbf{Sensor} & \textbf{Threshold} & \textbf{Mean Average accuracy} & \textbf{Maximum Average Accuracy} \\
\midrule
& 0.001 & 63.4\% & 97.1\% \\
& 0.01  & 63.5\% & 98.3\% \\
\textbf{Impeller} & 0.05  & 65.7\% & 96.1\% \\
& 0.1   & 70.0\% & \textbf{98.9}\% \\
& 0.2   & \textbf{74.4}\% & 94.9\% \\
\midrule
& 0.001 & 59.7\% & 91.0\% \\
& 0.01  & 61.0\% & 93.0\% \\
\textbf{Bearing Axial} & 0.05  & 64.8\% & 92.7\% \\
& 0.1   & 69.0\% & 94.5\% \\
& 0.2   & 74.1\% & 91.8\% \\
\bottomrule
\end{tabularx}
\end{minipage}
\end{table*}

\subsection{Limitations and Future Works}

\textcolor{black}{This study demonstrates the viability of encapsulated impeller sensors for pump monitoring but faces several limitations. First, experiments were conducted in a controlled laboratory environment using hydraulic oil under stable temperatures (25\degree C) and shallow submergence (1m), which may not reflect real-world conditions with crude oil, particulates, or extreme temperature/pressure fluctuations. Second, manual valve adjustments introduced timing inconsistencies ($\pm$~3s) during anomaly simulations, affecting model precision and potentially corrupting dataset labels. Third, the study focused on flow restriction and oil-level anomalies, omitting critical mechanical faults such as impeller imbalance or bearing wear. Finally, results are derived from a single vertical pump design, limiting generalization to horizontal or multi-stage configurations.}

\textcolor{black}{This intelligent encapsulated sensing system can be extended to other sensor types---such as temperature and pressure sensors---for condition monitoring in harsh environments. Further work on multi-sensor fusion integrating multi-mode sensor data could improve fault detection accuracy and robustness, and extend to root-cause diagnosis. Additionally, future research should prioritize field validation in operational oil wells with diverse pump designs and crude oil compositions. Automated anomaly simulation systems could eliminate manual valve variability, while expanded fault libraries should incorporate mechanical degradation modes (e.g., cavitation, shaft misalignment).  These steps will advance toward reliable, field-deployable impeller-based monitoring systems capable of reducing unplanned downtime in oil/gas operations.}

\subsection{Conclusion}

This study highlights the benefits of employing encapsulated accelerometers mounted directly on the impeller's bowl of a submerged vertical turbine pump for condition monitoring. Extensive testing under both normal- and anomaly-induced conditions demonstrated that the impeller-mounted accelerometer provides significantly richer diagnostic information compared to sensors positioned on the motor or bearings. Encapsulation proved to be a highly effective solution for protecting the sensor in harsh downhole environments without sacrificing its sensitivity---an essential capability for detecting early-stage failures in oil and gas applications. ML models further validated that data from the impeller-mounted sensor enhances predictive accuracy, reinforcing its critical role in advanced condition monitoring systems.
To the best of our knowledge, this research represents the first successful deployment of an encapsulated accelerometer directly on a submerged pump impeller, precisely at the defect-prone location, which in real-world applications can be situated up to hundreds of meters underground. While in the laboratory setting, sensors at other locations of the pump may provide nearly equally sufficient diagnostic information as the impeller-mounted sensor, on a real oil pump, an encapsulated, high-temperature and high-pressure-resilient, impeller-mounted sensor may be the only useful sensor. This achievement lays the groundwork for the development of improved downhole monitoring systems, offering a pathway toward more accurate and timely fault detection. By enabling enhanced operational reliability, this innovative approach has the potential to make a significant impact in the oil and gas industry.


\section{Experimental tests and methods}

\subsection{Experimental Setup}


\textcolor{black}{A laboratory-scale vertical turbine pump system was constructed to replicate oilfield operating conditions. The setup comprised:}

\begin{itemize}
    \item \textbf{Pump Assembly}: A 2-stage ITT Goulds VS6 vertical turbine pump with a rated speed of 1770 RPM and best efficiency point (BEP) at 62\% flow (26.2 US gallons per minute).
    \item \textbf{Drive System}: A 3 hp motor coupled to a variable frequency drive (VFD) for precise speed control.
    \item \textbf{Hydraulic Loop}: A closed-cycle oil circuit using MAG 1 AW ISO 46 hydraulic oil (Grainger, USA) circulated through a 100-gallon reservoir.
\end{itemize}

\textcolor{black}{The pump operated in a closed-loop configuration, drawing oil from and discharging to the same reservoir to maintain consistent fluid properties during testing.}

\subsubsection{Test Devices}


\textcolor{black}{The vibration signal acquisition system comprised:}

\begin{itemize}
    \item \textbf{Sensors}: Nine single-axis accelerometers (PCB Electronics Model 625B01, sensitivity: 100 mV/g) and one integrated cable accelerometer (CTC USA Model MEB360), mounted on the motor-pump assembly per API 610 specifications.
    \item \textbf{Signal Conditioning}: PCB Electronics Model 483C15 8-channel signal conditioner for amplification and noise filtering.
    \item \textbf{Data Acquisition}: National Instruments (NI) Model 781003-01 DAQ hardware interfaced with a custom LabVIEW-based software platform for real-time signal capture.
    \item \textbf{Computational Backbone}: Dedicated workstation for data storage and processing, sampling at 4000 Hz.
\end{itemize}

\textcolor{black}{This configuration ensured synchronized multi-channel vibration capture across all eight sensor locations (Table~\ref{tab:accel_locations}), with raw signals stored in 16-bit resolution for subsequent analysis.}

\subsubsection{Mounting Position}

\begin{figure}
    \centering
    \includegraphics[width=0.75\linewidth]{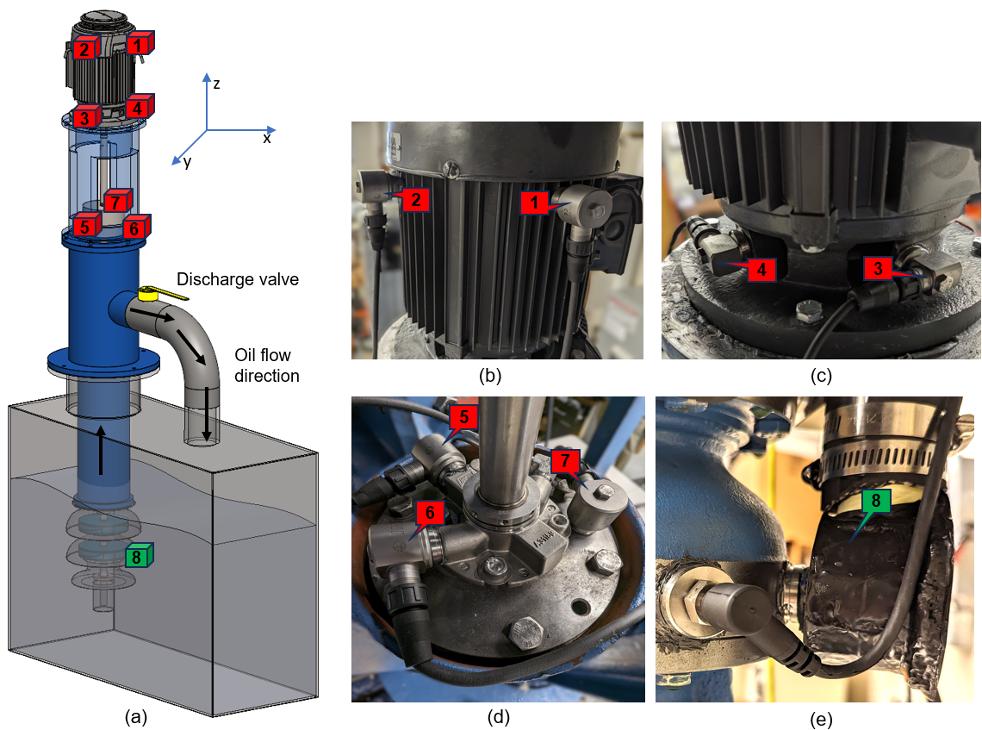}
    \caption{\centering Accelerometer mounting positions and directions. (a) CAD illustration of accelerometer mounting position and direction. Lab setup of (b) accelerometers mounted on motor outboard (c) accelerometers mounted on motor inboard (d) accelerometers mounted on bearings (e) encapsulated accelerometer on the impeller bowl. }
    \label{fig:accelerometer_mounting}
\end{figure}

\textcolor{black}{Accelerometers were mounted per API 610 guidelines, with encapsulated/non-encapsulated sensors on the impeller bowl (Figure \ref{fig:accelerometer_mounting}). Locations included axial/radial motor and bearing positions (Table \ref{tab:accel_locations}).}  

\newcolumntype{L}[1]{>{\raggedright\arraybackslash}p{#1}}
\newcolumntype{C}[1]{>{\centering\arraybackslash}p{#1}}
\newcolumntype{R}[1]{>{\raggedleft\arraybackslash}p{#1}}
\begin{table*}[h]
\caption{Accelerometer Sensor Locations}
\label{tab:accel_locations}
\centering
\begin{minipage}{1\linewidth}
\renewcommand{\thempfootnote}{\arabic{mpfootnote}}
\renewcommand{\arraystretch}{1.3}
\setlength{\tabcolsep}{5pt}
\begin{tabularx}{\textwidth}{C{2cm} C{6cm} C{3cm} Y}
\toprule
\textbf{Acc ID} & \textbf{Location} & \textbf{Axis} & \textbf{Description} \\
\midrule
&\textbf{Motor}&& \\
1  & Motor outboard MOB-0       & x & Parallel to the pump discharge piping \\
2  & Motor outboard MOB-90      & y & Perpendicular to the pump discharge piping \\
3  & Motor inboard MIB-0        & x & Parallel to the pump discharge piping \\
4  & Motor inboard MIB-90       & y & Perpendicular to the pump discharge piping \\
\midrule
&\textbf{Bearing}&&\\
5  & Pump Inboard PIB-0         & x & Parallel to the pump discharge piping \\
6  & Pump Inboard PIB-90        & y & Perpendicular to the pump discharge piping \\
7  & Pump Inboard PIB-A         & z & Axial to the flow \\
\midrule
&\textbf{Impeller's bowl}&& \\
8a & Encapsulated accelerometer  & x & Parallel to the pump discharge piping \\
8b & Non-encapsulated accelerometer & x & Parallel to the pump discharge piping \\
8c & Non-encapsulated accelerometer & y & Perpendicular to the pump discharge piping \\
\bottomrule
\end{tabularx}
\end{minipage}
\end{table*}

\subsubsection{Vibration Test Method}


\textcolor{black}{Vibration data were collected from eight accelerometers mounted on the vertical pump, including seven sensors above the oil level on the motor and bearings (Table~\ref{tab:accel_locations}) and one encapsulated accelerometer (Sr\#8a) on the submerged impeller bowl. The encapsulated sensor was protected by an epoxy-fluoroelastomer casing to withstand oil immersion, while a non-encapsulated control sensor (Sr\#8b) was installed adjacent to Sr\#8a for performance comparison. All sensors operated in hydraulic oil, with data synchronized at 4000 Hz. Although the laboratory setup permitted temporary use of non-encapsulated sensors, field conditions necessitate encapsulation to endure high-pressure crude oil and abrasive particulates.}


\textcolor{black}{The accelerometers were aligned both parallel and perpendicular to the pump's discharge pipe. Baseline vibration data were acquired under normal operating conditions: discharge valve fully open (100\%), motor speed fixed at 1770 RPM, and oil level maintained above the minimum submergence level (MSL) for two minutes.}

\textcolor{black}{Anomalies were simulated through two methodologies: (1) incremental valve closures (25\%, 50\%, 75\%) executed both gradually (over 30 seconds) and abruptly (<2 seconds), while maintaining constant motor speed; and (2) controlled oil level reductions to create sub-MSL and sub-impeller conditions.}

\textcolor{black}{Valve adjustments were performed manually due to equipment constraints, introducing minor timing inconsistencies ($\pm2$ seconds). However, post-hoc analysis shows these variations had negligible impact on model performance comparisons.}


\subsection{Anomaly Detection Experiment design}

To evaluate the effectiveness of a below-submergence-level sensor in detecting oil pump anomalies and to assess the impact of the proposed sensor encapsulation, we conducted a two-part experimental study.
In the first part, we compared the signals from a non-encapsulated impeller-mounted sensor with those from seven above-submergence-level sensors to assess their respective capabilities in representing the real-time pump head condition. ML models were trained on each sensor's normal signals and evaluated on their ability to detect anomalies using anomalous signals. Normal signals were recorded under standard operating conditions, while anomalous signals were collected during the previously described simulated anomalies.

\begin{table*}[h]
\caption{Datasets obtained from the experiments.}
\label{tab:datasets_experiments}
\centering
\begin{minipage}{1\linewidth}
\renewcommand{\thempfootnote}{\arabic{mpfootnote}}
\renewcommand{\arraystretch}{1.5}
\setlength{\tabcolsep}{5pt}
\begin{tabularx}{\textwidth}{L{8cm}C{4cm}C{2.5cm}C{2.5cm}}
\toprule
\textbf{Experiment Data} & \textbf{Data Composition} & \textbf{Abbreviation} & \textbf{\# of experiments} \\
\midrule
Abrupt Change -- Valve Closure & Normal + Abnormal & -- & 3\footnote{@25\%, 50\%, 75\% valve closure respectively} \\
Gradual Change -- Valve Closure @75\% & Normal + Abnormal & -- & 1 \\
Level Change -- Minimum Submergence Level & Normal & MSL & 1 \\
Level Change -- Below Minimum Submergence Level & Abnormal & BMSL & 1 \\
Level Change -- No Flow & Abnormal & -- & 1 \\
Valve Constant Flow & Normal & -- & 1 \\
\bottomrule
\end{tabularx}
\end{minipage}
\end{table*}

In the second part, we investigated the impact of encapsulation by comparing the signals from encapsulated and non-encapsulated impeller-mounted sensors. ML models were trained on signals from both types of sensors, and their anomaly detection performances were compared to assess differences in sensitivity and accuracy.
The pump's recommended operating speed of 1770 RPM was maintained throughout the experiments. To handle the time-series nature of the sensor data, the signals were segmented into smaller intervals, referred to as data samples, each lasting several hundred milliseconds to a few seconds. The data samples were processed using two approaches:
	Feature Extraction: Statistical, spectral, and temporal features were extracted from each data sample to create feature vectors suitable for ML models such as K-Nearest Neighbors (KNN), Deep Support Vector Data Description (DeepSVDD), and Autoencoder.
	Direct Signal Utilization: Normalized signals were directly utilized for models tailored to time-series analysis, including Long Short-Term Memory networks (LSTM) and Transformer-based architectures.
Details of data preprocessing techniques, model specifications, and evaluation criteria are provided in subsequent sections.

\subsubsection{Preprocessing}
Prior to any other preprocessing, the raw signals were standardized along each channel (sensor) to have a mean of 0 and a standard deviation of 1. This standardization ensures that the signals are brought into a uniform range, enhancing the robustness of ML models trained on the data. Notably, the standardization parameters (mean and standard deviation) were calculated from the training data and applied consistently to both training and testing datasets.
After standardization, the signals were partitioned into smaller data samples of varying durations $\tau$  and overlaps\ o\ . The partitioning used a sliding window approach, where the number of data points in each window was determined by the sampling frequency $\nu$, and $\tau\cdot\nu$ is the number of datapoints in each sliding window. There are $\frac{T}{\left(1-o\right)\cdot\tau\cdot\nu}$partitioned sliding windows for each signal sample. Specifically, we separate the normal signals from the abnormal signals in experiments where manual valve changes occurred in the middle, so that no windows contain mixed normal and abnormal datapoints. While this approach does not fully represent real-world scenarios where mixed data points might occur, it circumvents challenges associated with assigning individual anomaly scores to each data point in a window. Instead, all points in a window were assigned the label corresponding to the most anomalous point, a method often criticized for favoring random classifiers.[34] For oil pump operations, where failure detection within a few seconds is sufficient, this approach is justified.
We used $\tau = 0.25$ seconds, 0.5 seconds, 1 second, 3 seconds, $o$ = 0.9, 0.75, 0.5, 0 in our experiments to evaluate the overall performance of the models. The same configurations were applied to testing data to ensure consistency. Features extracted from normal and abnormal signal windows were normalized using the mean and standard deviation derived from the normal data used to train the models. Each sliding window of size $N\times8\ $(where N represents the number of data points and 8 is the number of sensors) was further divided along the sensor dimension to obtain ten different windows: eight $N\times1$ windows of each individual sensor, one $N\times7$ window of all sensors above oil level, and the original $N\times8$ windows encompassing all sensors. Models were trained on each of these datasets separately to quantitatively compare the informativeness of each sensor configuration.
Finally, a five-fold cross-validation (CV) was employed for all training and testing. The dataset consisted of six normal signal segments (detailed in Table \ref{tab:datasets_experiments}) and four abnormal segments for each encapsulation scheme. For the first fold, the normal parts of the abrupt change experiments at 25\% and 50\% valve closure were selected to combine with the four abnormal segments as testing data. In every other fold, one distinct segment from the rest of the normal data was selected to be the testing data. The normal segments not selected were used for training. The results were averaged and/or taken the maximum across all five folds to ensure robust performance evaluation.

\subsubsection{Feature extraction}

For each data sample and each channel, we extract 2 sets of features: the first consisting of 13 statistical features, and the second consisting of 15 spectral features. The features are concatenated in the same order as the channels, resulting in a $1\times8\cdot n_{feature}$feature vector per data sample. For each sample, this will produce $W\times8\cdot n_{feature}$, where $W$\ denotes the number of data samples extracted this way.
For models trained directly on the sliding windows of the standardized signals, we skipped this feature extraction process. However, because the sampling frequency of the sensor, 4000, is very high, and therefore the lengths of the sliding windows obtained on the scale of seconds are too long for almost any ML models. We tackle this problem by applying a 1-d gaussian filter along each channel of the signals with $\sigma=\frac{100}{6},\frac{500}{6}$, and subsampling the signals with step size 100 and 500. This is roughly equivalent to subsampling with a step size S\ and then applying a Gaussian weighted average with 99.7\% of its kernel mass falling inside each subsampling interval. Due to the computational costs, we only applied this one dimensionality reduction technique.

\subsubsection{Metrics}
Classifying unseen data points into normal or abnormal categories in unsupervised or semi-supervised settings is inherently challenging due to the absence of labeled anomalies during training. In these cases, models do not predict discrete labels directly but instead compute an "anomaly score" based on the data they are trained on. To evaluate the performance of our models in such settings, we emphasized metrics that address significant class imbalance and robustly reflect anomaly detection capabilities. 1) AUC-ROC score. The AUC-ROC score was used as a primary evaluation metric. It is resilient to class distribution shifts[40], making it particularly suitable for imbalanced datasets, such as those encountered in this study. 2) Average (Class) Accuracy. To further mitigate the impact of class imbalance, we calculated the average accuracy for both normal and abnormal classes. This metric ensures a balanced evaluation across the two classes. It is worth noting that we choose not to present the AUC-PR and F1-score since 80\% of normal data was used for training, with only 20\% included in the testing dataset alongside the anomalous samples (as shown in Table \ref{tab:datasets_experiments}). In our experiments, $\zeta$ for anomaly scores in each model is determined by sorting the scores predicted for all training data and selecting the score at a specific top percentile $k$. We tested five different $\zeta$s: 0.1\%, 1\%, 5\%, 10\%, and 20\%. Data collected after the simulated anomalies (at the one-minute mark) were labeled as anomalous for evaluation purposes.
In summary, we employed two sets of features---statistical and spectral---extracted from raw data using sliding windows. These sliding windows were configured with four durations (0.5, 1, 3, and 5 seconds) and four overlaps (0.9, 0.75, 0.5, and 0), resulting in 16 different configurations. Feature-based models were trained on these extracted feature datasets using selected hyperparameters and evaluated across all feature configurations at the five threshold levels to compute F1 scores.
For non-feature-based models, we skipped the feature extraction process and instead applied Gaussian moving average with subsampling step sizes of 100 and 500. This approach allowed us to directly process raw signals and compare performance with feature-based methods.


\medskip
\textbf{Acknowledgements} \par 

We wish to express our profound appreciation to the Saudi Arabian Oil Company (Saudi Aramco) for their generous financial support and expert technical guidance. Their collaboration has been instrumental in identifying and addressing a critical technical challenge that is of paramount importance to the industry. Our research provides innovative solutions to this pressing issue, and their support has greatly contributed to our endeavor. We would also like to extend our gratitude to the Massachusetts Institute of Technology (MIT) for technical and engineering insights and guidance, which further enriched the quality and depth of our work.

\medskip
\textbf{CONFLICT OF INTEREST} \par 
The authors declare no conflict of interest. This work is supported by the Saudi Arabian Oil Company (Saudi Aramco). Any opinions, findings, conclusions, or recommendations expressed in this material are those of the authors and do not necessarily reflect the views of Saudi Aramco.

\medskip

%
\bibliography{ref}

\begin{thebibliography}{10}

\bibitem{corley1980vibration}
J.~Corley, ``Vibration problems of large vertical pumps and motors,'' 1980.

\bibitem{lakal2022sensing}
N.~Lakal, A.~H. Shehri, K.~W. Brashler, S.~P. Wankhede, J.~Morse, and X.~Du, ``Sensing technologies for condition monitoring of oil pump in harsh environment,'' {\em Sensors and Actuators A: Physical}, vol.~346, p.~113864, 2022.

\bibitem{wankhede2022encapsulating}
S.~Wankhede, X.~Du, A.~Alshehri, K.~Brashler, and D.~Turcan, ``Encapsulating and inkjet-printing electronics on flexible substrates for harsh environment,'' in {\em International Manufacturing Science and Engineering Conference}, vol.~86601, p.~V001T03A001, American Society of Mechanical Engineers, 2022.

\bibitem{wankhede2023encapsulating}
S.~P. Wankhede, X.~Du, K.~W. Brashler, M.~M. Ba'adani, D.~C. Turcan, A.~H. Shehri, and K.~Youcef-Toumi, ``Encapsulating commercial accelerometers with epoxy and fluoroelastomer for harsh hydrocarbon fluid environment,'' {\em Scientific Reports}, vol.~13, no.~1, p.~19815, 2023.

\bibitem{wankhede2023chem}
S.~P. Wankhede, A.~H. Alshehri, and X.~Du, ``Encapsulating and inkjet-printing flexible conductive patterns on a fluoroelastomer for harsh hydrocarbon fluid environments,'' {\em Journal of Materials Chemistry C}, vol.~11, no.~12, pp.~3964--3980, 2023.

\bibitem{zhang2022vibration}
Y.~Zhang, J.~Liu, X.~Yang, H.~Li, S.~Chen, W.~Lv, W.~Xu, J.~Zheng, and D.~Wang, ``Vibration analysis of a high-pressure multistage centrifugal pump,'' {\em Scientific Reports}, vol.~12, no.~1, p.~20293, 2022.

\bibitem{bai2019vibration}
L.~Bai, L.~Zhou, X.~Jiang, Q.~Pang, and D.~Ye, ``Vibration in a multistage centrifugal pump under varied conditions,'' {\em Shock and Vibration}, vol.~2019, no.~1, p.~2057031, 2019.

\bibitem{Wang2018}
K.~Wang, C.~Wang, C.~Xia, H.~Liu, and Z.~Zhang, ``Experimental measurement of cavitation-induced vibration characteristics in a multi-stage centrifugal pump,'' {\em Journal of Chemical Engineering of Japan}, vol.~51, pp.~203--209, Mar 2018.

\bibitem{Leader1985}
M.~E. Leader, ``A solution for variable speed vertical pump vibration problems,'' lecture notes, pump symposium, Turbomachinery Laboratories, Department of Mechanical Engineering, Texas A\&M University, 1985.

\bibitem{Smith1986}
D.~R. Smith and G.~M. Woodward, ``Vibration analysis of vertical pumps,'' tech. rep., Texas A\&M University, Turbomachinery Laboratories, 1986.
\newblock Pump Symposium Lecture.

\bibitem{Schiavello2004}
B.~Schiavello, D.~R. Smith, and S.~M. Price, ``Abnormal vertical pump suction recirculation problems due to pump-system interaction,'' tech. rep., Engineering Dynamics Incorporated, 2004.
\newblock Technical report; EDI Publication.

\bibitem{Le_TurbinePump}
T.~Le, {\em Machinery Health Monitoring System for Vertical Turbine Pump}.
\newblock PhD thesis, -, -.
\newblock Ph.D. thesis; exact year and institution not found.

\bibitem{Shahrooi2009}
S.~Shahrooi, I.~H. Metselaar, Z.~Huda, and M.~Asayesh, ``Numerical and experimental vibration analysis for fatigue failure investigation in a vertical axis pump station,'' in {\em Proceedings of the Hong Kong Pump Symposium}, (Hong Kong), 2009.

\bibitem{AbdelRahman2009}
S.~M. Abdel-Rahman and S.~A.~A. El-Shaikh, ``Diagnosis vibration problems of pumping stations: Case studies,'' tech. rep., -, 2009.
\newblock Case-study report; publisher not specified.

\bibitem{LeaderConnerLucas}
M.~E. Leader, K.~J. Conner, and J.~D. Lucas, ``Evaluating and correcting subsynchronous vibration in vertical pumps,'' tech. rep., -, -.
\newblock Report; year not specified.

\bibitem{Fetyan2014}
K.~Fetyan and D.~El-Gazzar, ``Effect of motor vibration problem on the power quality of water pumping stations,'' {\em Water Science}, vol.~28, pp.~31--41, 2014.

\bibitem{ElGazzar2017}
D.~M. El-Gazzar, ``Finite element analysis for structural modification and control resonance of a vertical pump,'' {\em Alexandria Engineering Journal}, vol.~56, pp.~695--707, 2017.

\bibitem{Stan2018}
M.~Stan, I.~Pana, M.~Minescu, A.~Ichim, and C.~Teodoriu, ``Centrifugal pump monitoring and determination of pump characteristic curves using experimental and analytical solutions,'' {\em Processes}, vol.~6, p.~18, 2018.

\bibitem{AlTobi2019}
M.~A.~S. AlTobi, G.~Bevan, P.~Wallace, D.~Harrison, and K.~P. Ramachandran, ``Fault diagnosis of a centrifugal pump using mlp-gabp and svm with cwt,'' {\em Engineering Science and Technology, International Journal}, vol.~22, pp.~854--861, 2019.

\bibitem{ADBench}
S.~Han, X.~Hu, H.~Huang, M.~Jiang, and Y.~Zhao, ``Adbench: Anomaly detection benchmark,'' {\em Advances in neural information processing systems}, vol.~35, pp.~32142--32159, 2022.

\bibitem{revisitingTSOD}
K.-H. Lai, D.~Zha, J.~Xu, Y.~Zhao, G.~Wang, and X.~Hu, ``Revisiting time series outlier detection: Definitions and benchmarks,'' in {\em Thirty-fifth conference on neural information processing systems datasets and benchmarks track (round 1)}, -.

\bibitem{noFreeLunchTheorem}
D.~H. Wolpert and W.~G. Macready, ``No free lunch theorems for optimization,'' {\em IEEE transactions on evolutionary computation}, vol.~1, no.~1, pp.~67--82, 1997.

\bibitem{Lai2021RevisitingTSOD}
K.-H. Lai, D.~Zha, J.~Xu, Y.~Zhao, G.~Wang, and X.~Hu, ``Revisiting time series outlier detection: Definitions and benchmarks,'' in {\em Thirty-fifth conference on neural information processing systems datasets and benchmarks track (round 1)}, 2021.

\bibitem{Liu2008IForest}
F.~T. Liu, K.~M. Ting, and Z.-H. Zhou, ``Isolation forest,'' in {\em 2008 Eighth IEEE International Conference on Data Mining}, pp.~413--422, 2008.

\bibitem{Fix1989(KNN)}
E.~Fix and J.~L. Hodges, ``Discriminatory analysis. nonparametric discrimination: Consistency properties,'' {\em International Statistical Review}, vol.~57, pp.~238--247, 1989.

\bibitem{He2003}
Z.~He, X.~Xu, and S.~Deng, ``Discovering cluster-based local outliers,'' {\em Pattern Recognition Letters}, vol.~24, pp.~1641--1650, 2003.

\bibitem{Li2020COPOD}
Z.~Li, Y.~Zhao, N.~Botta, C.~Ionescu, and X.~Hu, ``Copod: Copula-based outlier detection,'' in {\em 2020 IEEE International Conference on Data Mining (ICDM)}, pp.~1118--1123, 2020.

\bibitem{Rumelhart1986}
D.~E. Rumelhart, G.~E. Hinton, and R.~J. Williams, ``Learning internal representations by error propagation,'' in {\em Parallel Distributed Processing: Explorations in the Microstructure of Cognition, Vol. I} (D.~E. Rumelhart and J.~L. McClelland, eds.), pp.~318--362, Cambridge, MA: MIT Press, 1986.

\bibitem{Ruff2018}
L.~Ruff {\em et~al.}, ``Deep one-class classification,'' in {\em Proceedings of the 35th International Conference on Machine Learning}, pp.~4393--4402, PMLR, 2018.

\bibitem{Malhotra2016}
P.~Malhotra {\em et~al.}, ``Lstm-based encoder-decoder for multi-sensor anomaly detection.'' \url{https://doi.org/10.48550/arXiv.1607.00148}, 2016.

\bibitem{wagner2023timesead}
D.~Wagner, T.~Michels, F.~C. Schulz, A.~Nair, M.~Rudolph, and M.~Kloft, ``Timesead: Benchmarking deep multivariate time-series anomaly detection,'' {\em Transactions on Machine Learning Research}, -.

\bibitem{Schmidl2022}
S.~Schmidl, P.~Wenig, and T.~Papenbrock, ``Anomaly detection in time series: A comprehensive evaluation,'' {\em Proceedings of the VLDB Endowment}, vol.~15, pp.~1779--1797, 2022.

\bibitem{HeZhao2019}
Y.~He and J.~Zhao, ``Temporal convolutional networks for anomaly detection in time series,'' {\em Journal of Physics: Conference Series}, vol.~1213, p.~042050, 2019.

\bibitem{Shen2020}
L.~Shen, Z.~Li, and J.~T. Kwok, ``Time series anomaly detection using temporal hierarchical one-class network,'' in {\em Advances in Neural Information Processing Systems, vol. 33}, pp.~13016--13026, Curran Associates, Inc., 2020.

\bibitem{Xu2022}
J.~Xu, H.~Wu, J.~Wang, and M.~Long, ``Anomaly transformer: Time series anomaly detection with association discrepancy.'' \url{https://doi.org/10.48550/arXiv.2110.02642}, 2022.

\bibitem{Kim2022}
S.~Kim, K.~Choi, H.-S. Choi, B.~Lee, and S.~Yoon, ``Towards a rigorous evaluation of time-series anomaly detection,'' in {\em Proceedings of the AAAI Conference on Artificial Intelligence}, vol.~36, pp.~7194--7201, 2022.

\bibitem{ZamanzadehDarban2024}
Z.~Zamanzadeh~Darban, G.~I. Webb, S.~J. Pan, C.~Aggarwal, and M.~Salehi, ``Deep learning for time series anomaly detection: A survey,'' {\em ACM Computing Surveys}, vol.~57, no.~15, pp.~1--42, 2024.

\bibitem{Choi2021}
K.~Choi, J.~Yi, C.~Park, and S.~Yoon, ``Deep learning for anomaly detection in time-series data: Review, analysis, and guidelines,'' {\em IEEE Access}, vol.~9, pp.~120043--120065, 2021.

\bibitem{Zhang2024}
K.~Zhang {\em et~al.}, ``Self-supervised learning for time series analysis: Taxonomy, progress, and prospects,'' {\em IEEE Transactions on Pattern Analysis and Machine Intelligence}, vol.~46, pp.~6775--6794, 2024.

\end{thebibliography}







\newpage
\text{\centering Graphical Abstract of the manuscript}

\begin{figure}[h]
    \centering
    \includegraphics[width=1\linewidth]{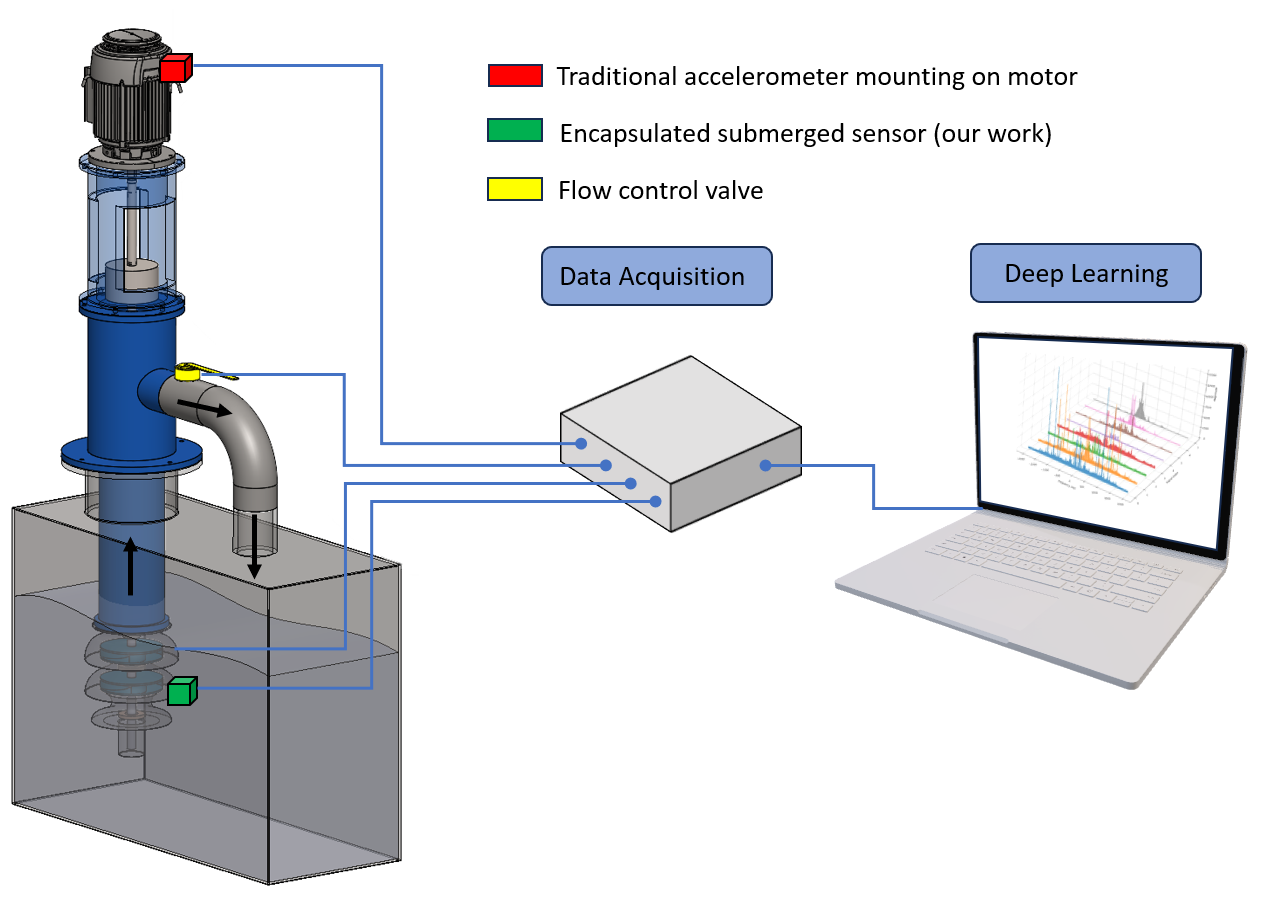}
    \label{fig:graphical_abstract}
\end{figure}

In the oil and gas industry, a novel condition monitoring approach places encapsulated accelerometers directly on pump impellers submerged in oil wells. These robust sensors survive harsh oil well conditions and use deep learning to detect faults in real-time. This first-of-its-kind intelligent in-situ monitoring system achieves 98.0\% fault detection accuracy (AUC=0.987), outperforming traditional above-ground sensors by 5.3\%.

\newpage

{\centering \section{Supplementary}}

\setcounter{figure}{0}
\renewcommand{\figurename}{Figure}
\renewcommand{\thefigure}{S\arabic{figure}}

\setcounter{table}{0}
\renewcommand{\tablename}{Table}
\renewcommand{\thetable}{S\arabic{table}}

\begin{figure}[h]
    \centering
    \includegraphics[width=1\linewidth]{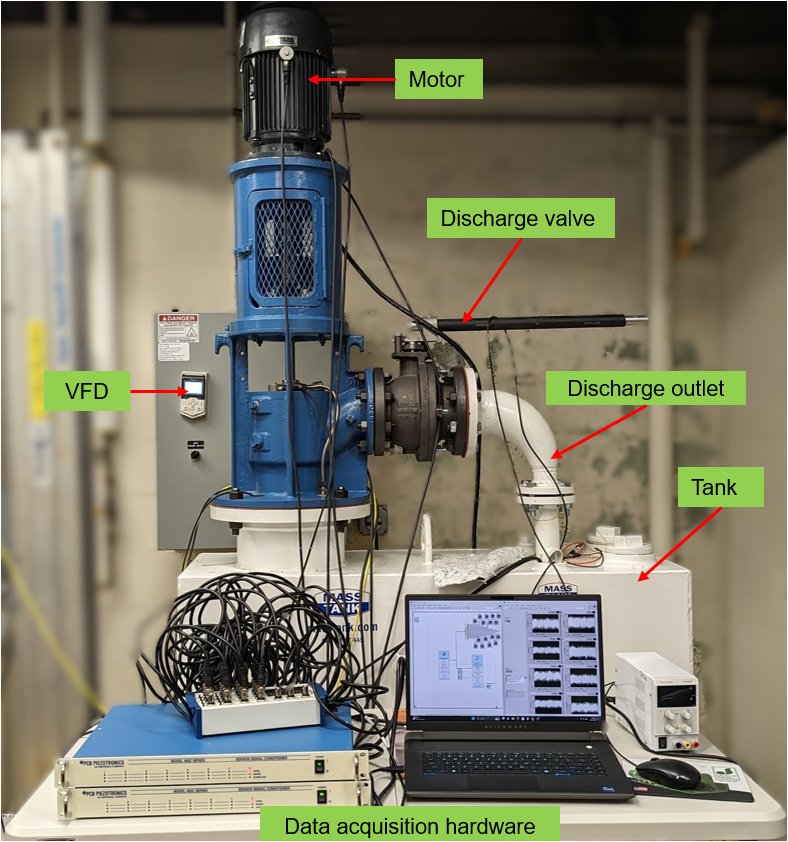}
    \caption{\centering Oil Pump experimental setup}
    \label{figs:oil_pump_experimental_setup}
\end{figure}

\subsection{Test Cases}

\textbf{Case 1: Normal conditions}

The pump motor speed was kept constant at 1770 RPM for 2 minutes with the discharge valve fully open. (100\% open)

\textbf{Case 2: Abnormal conditions by changing the discharge valve position}

\begin{enumerate}
    \item Abrupt Valve Closure
    \begin{enumerate}
        \item 
        The pump motor speed was kept constant at 1770 RPM for 1 minute with the discharge valve 100\% open and a sudden closure of the discharge valve to 25\% for 1 minute.
        \item 
        The pump motor speed was kept constant at 1770 RPM for 1 minute with the discharge valve 100\% open and a sudden closure of the discharge valve to 50\% for 1 minute.
        \item 
        The pump motor speed was kept constant at 1770 RPM for 1 minute with the discharge valve 100\% open and a sudden closure of the discharge valve to 75\% for 1 minute. 
    \end{enumerate}

    \item Gradual Valve Closure
    \begin{enumerate}
        \item 
        The pump motor speed was kept constant at 1770 RPM for 1 minute with the discharge valve 100\% open followed by a gradual closure (manually) of the discharge valve to 75\% in 1 minute.
    \end{enumerate}

    \item Abrupt Valve Closure followed by Abrupt Valve Opening
    \begin{enumerate}
        \item 
        The pump motor speed was kept constant at 1770 RPM for 1 minute with the discharge valve 100\% open followed by a sudden closure of the discharge valve to 25\% for 0.5 minutes, followed by a sudden opening of the discharge valve to 100\% for 0.5 minutes.
        \item 
        The pump motor speed was kept constant at 1770 RPM for 1 minute with the discharge valve 100\% open followed by a sudden closure of the discharge valve to 50\% for 0.5 minutes, followed by a sudden opening of the discharge valve to 100\% for 0.5 minutes.
        \item 
        The pump motor speed was kept constant at 1770 RPM for 1 minute with the discharge valve 100\% open followed by a sudden closure of the discharge valve to 75\% for 0.5 minutes, followed by a sudden opening of the discharge valve to 100\% for 0.5 minutes.
    \end{enumerate}
\end{enumerate}

It should be noted that changes to the valve position were made manually, which may have resulted in errors in the data collection during the specified time frame.

\textbf{Case 3: Abnormal conditions by changing the oil level }

Vibration data was collected while reducing the oil level in two scenarios:
\begin{enumerate}
    \item Below MSL: The oil level in the tank was below MSL but there was still oil flow.
    \item Below the impeller: When the oil level in the tank is dropped below the stage 1 impeller, there is no oil flow.
\end{enumerate}

\newcolumntype{Y}{>{\RaggedRight\arraybackslash}X}

\begin{table*}[h]
\caption{Statistical features computed from a time series $\{x_i\}_{i=1}^N$.}
\label{tab:statistical_features}
\centering
\begin{minipage}{1\linewidth}
\renewcommand{\arraystretch}{2}
\setlength{\tabcolsep}{5pt}
\begin{tabularx}{\textwidth}{YY}
\toprule
\textbf{Feature} & \textbf{Definition} \\
\midrule
1. Absolute energy & $E = \sum_{i=1}^N x_i^2$ \\
\midrule
2. Average power & $P = \frac{1}{N}\sum_{i=1}^N x_i^2$ \\
\midrule
3. Entropy & $H = -\sum_{i=1}^N p_i \log(p_i), \quad p_i = \frac{|x_i|}{\sum_{j=1}^N |x_j|}$ \\
\midrule
4. Kurtosis & $K = \frac{\frac{1}{N}\sum_{i=1}^N (x_i-\bar x)^4}{(\frac{1}{N}\sum_{i=1}^N (x_i-\bar x)^2)^2}$ \\
\midrule
5. Max & $\max = \max(x_1, x_2, \dots, x_N)$ \\
\midrule
6. Mean & $\bar x = \frac{1}{N}\sum_{i=1}^N x_i$ \\
\midrule
7. Mean absolute deviation & $\mathrm{MAD} = \frac{1}{N}\sum_{i=1}^N |x_i - \bar x|$ \\
\midrule
8. Median & $\mathrm{median} = \text{middle value of sorted }\{x_1, x_2, \dots, x_N\}$ \\
\midrule
9. Median absolute deviation & $\mathrm{MAD}_{\mathrm{median}} = |x_i - \mathrm{median}(x)|,\; i=1,2,\dots,N$ \\
\midrule
10. Min & $\min = \min(x_1, x_2, \dots, x_N)$ \\
\midrule
11. Root mean square & $\mathrm{RMS} = \sqrt{\frac{1}{N}\sum_{i=1}^N x_i^2}$ \\
\midrule
12. Skewness & $S = \frac{\frac{1}{N}\sum_{i=1}^N (x_i-\bar x)^3}{(\frac{1}{N}\sum_{i=1}^N (x_i-\bar x)^2)^{3/2}}$ \\
\midrule
13. Standard deviation & $\sigma = \sqrt{\frac{1}{N}\sum_{i=1}^N (x_i-\bar x)^2}$ \\
\bottomrule
\end{tabularx}
\end{minipage}
\end{table*}

\begin{table*}[!h]
  \caption{Statistical features computed from a time series $\{x_i\}_{i=1}^N$.}
  \label{tab:spectral_features}
  \centering
  \begin{minipage}{1\linewidth}
    \renewcommand{\arraystretch}{3}
    \setlength{\tabcolsep}{5pt}
    \begin{tabularx}{\textwidth}{YY}
      \toprule
      \textbf{Feature} & \textbf{Definition} \\
      \midrule
      1. Max power spectrum & $\mathrm{MaxPower} = \max\bigl(|X(f)|^2\bigr),\quad f \in [0, f_s/2]$ \\
      \midrule
      2. Maximum frequency & $f_{\max} = \arg\max\bigl(|X(f)|^2\bigr),\quad f \in [0, f_s/2]$ \\
      \midrule
      3. Median frequency & $\displaystyle \int_0^{f^{\mathrm{median}}} |X(f)|^2\,df = \tfrac12 \int_0^{f_s/2}|X(f)|^2\,df$ \\
      \midrule
      4. Power bandwidth & $\displaystyle \int_0^{f^b}|X(f)|^2\,df = \epsilon \int_0^{f_s/2}|X(f)|^2\,df,\quad \epsilon=0.95$ \\
      \midrule
      5. Spectral centroid & $C = \frac{\sum_f f\cdot|X(f)|}{\sum_f |X(f)|}$ \\
      \midrule
      6. Spectral decrease & $D = \frac{\sum_{k=2}^K \frac{|X(k)|-|X(1)|}{k-1}}{\sum_{k=1}^K |X(k)|}$ \\
      \midrule
      7. Spectral distance & $\mathrm{Dist}=\sqrt{\sum_f\bigl(|X(f)|^2 - |Y(f)|^2\bigr)^2},$ where $Y$ is the fitted linear regression. \\
      \midrule
      8. Spectral entropy & $H_s = -\sum_f p(f)\log p(f),\quad p(f)=\frac{|X(f)|^2}{\sum_f|X(f)|^2}$ \\
      \midrule
      9. Spectral kurtosis & $K_s = \frac{\frac{1}{N}\sum_f(|X(f)|^2-\overline{P})^4}{(\frac{1}{N}\sum_f(|X(f)|^2-\overline{P})^2)^2},\quad \overline{P}=\frac{1}{N}\sum_f|X(f)|^2$ \\
      \midrule
      10. Spectral roll-off & $\displaystyle \int_0^{f^{\mathrm{roll-off}}}|X(f)|^2\,df = r\int_0^{f_s/2}|X(f)|^2\,df,\; r=0.95$ \\
      \midrule
      11. Spectral roll-on & $\displaystyle \int_{f^{\mathrm{roll-on}}}^{f_s/2}|X(f)|^2\,df = r\int_0^{f_s/2}|X(f)|^2\,df,\; r=0.05$ \\
      \midrule
      12. Spectral skewness & $S_s = \frac{\frac{1}{N}\sum_f(|X(f)|^2-\overline{P})^3}{(\frac{1}{N}\sum_f(|X(f)|^2-\overline{P})^2)^{3/2}}$ \\
      \midrule
      13. Spectral spread & $\mathrm{Spread} = \sqrt{\frac{\sum_f (f-C)^2|X(f)|}{\sum_f|X(f)|}}$ \\
      \midrule
      14. Spectral variation & $V = 1 - \frac{\sum_{k=1}^{N-1}A_k\cdot A_{k+1}}{\sqrt{(\sum_{k=1}^{N-1}A_k^2)\,(\sum_{k=1}^{N-1}A_{k+1}^2)}}$ \\
      \midrule
      15. Wavelet absolute mean & $M_w = \frac{1}{N}\sum_{i=1}^N |w_i|$ \\
      \bottomrule
    \end{tabularx}
  \end{minipage}
\end{table*}

*All feature computations were done with the Python package \textit{tsfel}.

\newcolumntype{Y}{>{\Centering\arraybackslash}X}

\begin{table*}[h]
\caption{Default hyperparameters used by the deep models}
\label{tab:default_hyperparameters}
\centering
\begin{minipage}{1\linewidth}
\renewcommand{\thempfootnote}{\arabic{mpfootnote}}
\renewcommand{\arraystretch}{1.5}
\setlength{\tabcolsep}{5pt}
\begin{tabularx}{\textwidth}{L{3cm}YYYY}
\toprule
\textbf{hyperparameters} & \textbf{Anomaly-Transformer} & \textbf{LSTM-Autoencoder} & \textbf{TCN-AD} & \textbf{THOC} \\
\midrule
hidden size        & 512 & 512 & 2 & 32 \\
\# of heads         & 8   & -   & - & -  \\
\# of layers        & 3   & 3   & 3 & $\log N + 1\,^*$\footnote{$N$ is the window size.} \\
dropout rate       & 0.0 & 0.0 & 0.0 & 0.0 \\
skip connections   & True & False & True & - \\
autoregressive     & False & - & False & - \\
kernel size        & - & - & 4 & - \\
\bottomrule
\end{tabularx}
\end{minipage}
\end{table*}

\begin{figure}[h]
    \centering
    \includegraphics[width=1\linewidth]{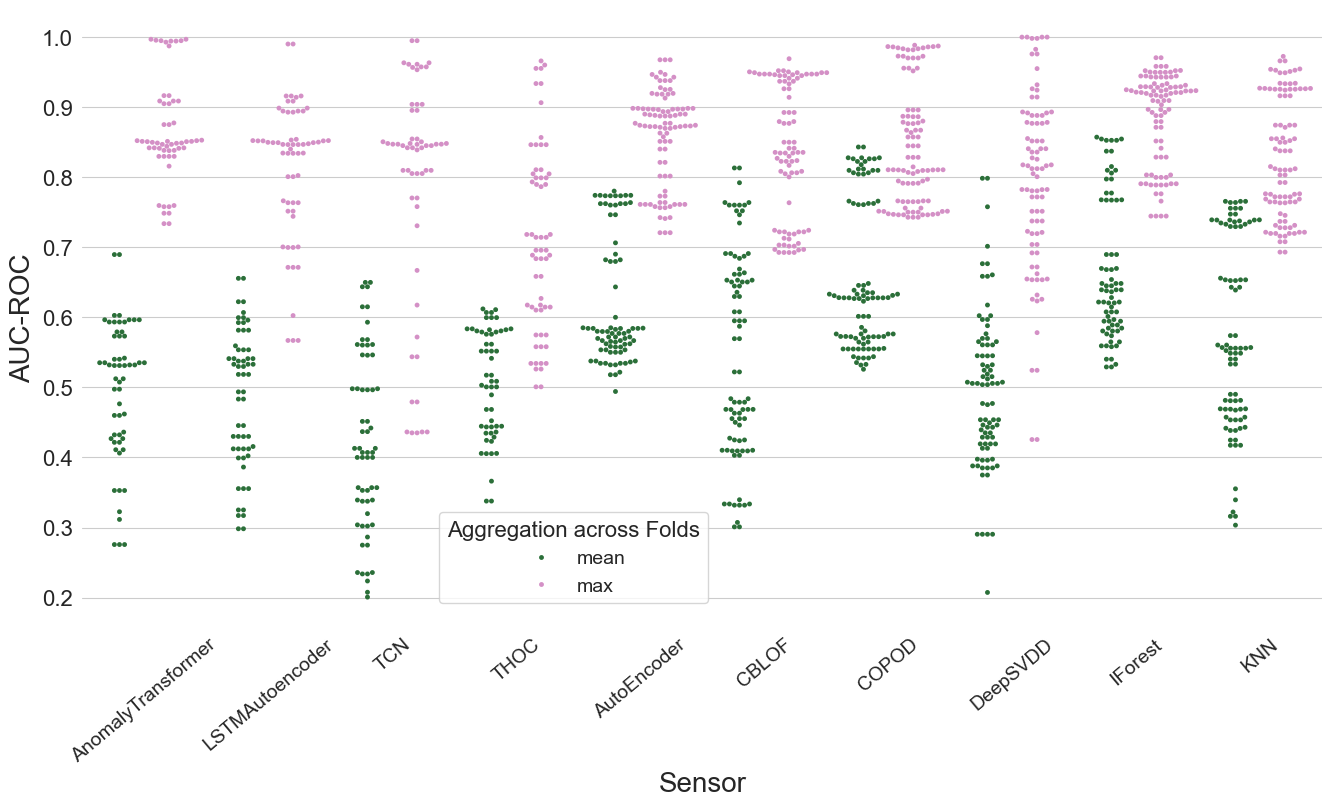}
    \caption{\centering Maximum vs Mean AUC-ROC of each model across CV folds, with dots representing results from a random subsample of experiments. }
    \label{fig:AUCROC_model_CVaggregation_swarm}
\end{figure}

\newcolumntype{Y}{>{\Centering\arraybackslash}X}

\begin{table*}[h]
\caption{AUC-ROC of Encapsulated Impeller Sensor and Other Sensors with one Standard Deviation}
\label{tabS:AUCROC_sensor_model}
\centering
\begin{minipage}{1\linewidth}
\renewcommand{\thempfootnote}{\arabic{mpfootnote}}
\renewcommand{\arraystretch}{1.5}
\setlength{\tabcolsep}{5pt}
\begin{tabularx}{\textwidth}{L{3cm}YYYYYYYY}

\toprule
\textbf{Model} & \textbf{Impeller (encapsulated)} & \textbf{Bearing pl} & \textbf{Bearing pp} & \textbf{Bearing ax} & \textbf{Motor inboard pl} & \textbf{Motor inboard pp} & \textbf{Motor outboard pl }& \textbf{Motor outboard pp }\\
\midrule
AnomalyTransformer & 
0.816 \(\pm\) 0.073 & 
\textbf{0.875} \(\pm\) 0.074\footnote{Best performing sensor for each model is highlighted in bold} & 
0.715 \(\pm\) 0.140 & 
0.807 \(\pm\) 0.087 & 
0.821 \(\pm\) 0.140 & 
0.748 \(\pm\) 0.056 & 
0.801 \(\pm\) 0.158 & 
0.781 \(\pm\) 0.195 \\

AutoEncoder & 
\textbf{0.820 \(\pm\) 0.066} & 
0.790 \(\pm\) 0.110 & 
0.738 \(\pm\) 0.082 & 
0.793 \(\pm\) 0.085 & 
0.772 \(\pm\) 0.062 & 
0.771 \(\pm\) 0.067 & 
0.729 \(\pm\) 0.104 & 
0.727 \(\pm\) 0.063 \\

CBLOF & 
\textbf{0.906 \(\pm\) 0.093} & 
0.655 \(\pm\) 0.092 & 
0.636 \(\pm\) 0.118 & 
0.605 \(\pm\) 0.126 & 
0.693 \(\pm\) 0.092 & 
0.717 \(\pm\) 0.056 & 
0.637 \(\pm\) 0.103 & 
0.662 \(\pm\) 0.083 \\

COPOD & 
\textbf{0.821 \(\pm\) 0.084} & 
0.765 \(\pm\) 0.109 & 
0.709 \(\pm\) 0.072 & 
0.787 \(\pm\) 0.060 & 
0.730 \(\pm\) 0.075 & 
0.711 \(\pm\) 0.064 & 
0.721 \(\pm\) 0.091 & 
0.682 \(\pm\) 0.077 \\

DeepSVDD & 
0.750 \(\pm\) 0.130 & 
\textbf{0.782 \(\pm\) 0.158} & 
0.754 \(\pm\) 0.156 & 
0.792 \(\pm\) 0.155 & 
0.706 \(\pm\) 0.130 & 
0.691 \(\pm\) 0.113 & 
0.684 \(\pm\) 0.135 & 
0.664 \(\pm\) 0.109 \\

IForest & 
\textbf{0.865 \(\pm\) 0.059} & 
0.771 \(\pm\) 0.107 & 
0.734 \(\pm\) 0.079 & 
0.778 \(\pm\) 0.079 & 
0.792 \(\pm\) 0.083 & 
0.789 \(\pm\) 0.063 & 
0.785 \(\pm\) 0.069 & 
0.760 \(\pm\) 0.061 \\

KNN & 
\textbf{0.889 \(\pm\) 0.087} & 
0.699 \(\pm\) 0.102 & 
0.655 \(\pm\) 0.120 & 
0.606 \(\pm\) 0.130 & 
0.728 \(\pm\) 0.077 & 
0.731 \(\pm\) 0.051 & 
0.632 \(\pm\) 0.076 & 
0.689 \(\pm\) 0.074 \\

LSTMAutoencoder & 
0.817 \(\pm\) 0.091 & 
\textbf{0.842 \(\pm\) 0.094} & 
0.700 \(\pm\) 0.079 & 
0.784 \(\pm\) 0.080 & 
0.819 \(\pm\) 0.152 & 
0.756 \(\pm\) 0.054 & 
0.812 \(\pm\) 0.202 & 
0.740 \(\pm\) 0.233 \\

TCN & 
0.793 \(\pm\) 0.156 & 
\textbf{0.819 \(\pm\) 0.151} & 
0.691 \(\pm\) 0.101 & 
0.736 \(\pm\) 0.114 & 
0.792 \(\pm\) 0.161 & 
0.697 \(\pm\) 0.079 & 
0.768 \(\pm\) 0.205 & 
0.721 \(\pm\) 0.209 \\

THOC & 
0.704 \(\pm\) 0.133 & 
\textbf{0.759 \(\pm\) 0.095} & 
0.606 \(\pm\) 0.076 & 
0.671 \(\pm\) 0.093 & 
0.611 \(\pm\) 0.090 & 
0.608 \(\pm\) 0.063 & 
0.618 \(\pm\) 0.067 & 
0.596 \(\pm\) 0.062 \\

\bottomrule
\end{tabularx}
\end{minipage}
\end{table*}

\end{document}